\documentclass[journal]{IEEEtran}
\usepackage{amsmath,amsfonts}
\usepackage{algorithmic}
\usepackage{algorithm}
\usepackage{array}
\usepackage[caption=false,font=normalsize,labelfont=sf,textfont=sf]{subfig}
\usepackage{textcomp}
\usepackage{stfloats}
\usepackage{url}
\usepackage{verbatim}
\usepackage{graphicx}
\usepackage{subfig}
\usepackage{cite}
\usepackage{xcolor}
\newcommand{\red}[1]{\textcolor{black}{#1}}
\begin{document}

\title{Exoplanet Imaging via Differentiable Rendering}

\author{Brandon Y. Feng, Rodrigo Ferrer-Chávez, Aviad Levis,\\ Jason J. Wang, Katherine L. Bouman, and William T. Freeman
        % <-this % stops a space
\thanks{B. Y. Feng and W. T. Freeman are with the Computer Science and Artificial Intelligence Laboratory, Massachusetts Institute of Technology (email: branfeng@mit.edu, billf@mit.edu). R. Ferrer-Chávez and J. J. Wang are with the Department of Physics \& Astronomy, Northwestern University (email: rodrigoferrer-chavez2029@u.northwestern.edu, jason.wang@northwestern.edu). A. Levis is with the Department of Computer Science, the University of Toronto (email: aviad.levis@gmail.com). 
K. L. Bouman is with the Departments of Computing and Mathematical Sciences, Electrical Engineering, and Astronomy, California Institute of Technology (email: klbouman@caltech.edu).}% <-this % stops a space
%\thanks{Manuscript received April 19, 2021; revised August 16, 2021.}
}

\markboth{IEEE TRANSACTIONS ON COMPUTATIONAL IMAGING}%
{}


\maketitle

\begin{abstract}
Direct imaging of exoplanets is crucial for advancing our understanding of planetary systems beyond our solar system, but it faces significant challenges due to the high contrast between host stars and their planets. 
Wavefront aberrations introduce speckles in the telescope science images, which are patterns of diffracted starlight that can mimic the appearance of planets, complicating the detection of faint exoplanet signals. 
Traditional post-processing methods, operating primarily in the image intensity domain, do not integrate wavefront sensing data.
These data, measured mainly for adaptive optics corrections, have been overlooked as a potential resource for post-processing, partly due to the challenge of the evolving nature of wavefront aberrations.
{In this paper, we present a differentiable rendering approach that leverages these wavefront sensing data to improve exoplanet detection}.
Our differentiable renderer models wave-based light propagation through a coronagraphic telescope system, allowing gradient-based optimization to significantly improve starlight subtraction and increase sensitivity to faint exoplanets. 
Simulation experiments based on the James Webb Space Telescope configuration demonstrate the effectiveness of our approach, achieving substantial improvements in contrast and planet detection limits. 
Our results showcase how the computational advancements enabled by differentiable rendering can revitalize previously underexploited wavefront data, opening new avenues for enhancing exoplanet imaging and characterization.
\end{abstract}


\section{Introduction}
\label{sec:intro}

Exoplanets, or extrasolar planets, are planets that orbit stars outside our solar system.
Studying exoplanets is crucial for understanding planetary system formation and evolution.
Moreover, it provides insights into the conditions necessary for habitability and the potential for finding life beyond Earth, addressing the profound question: ``Are we alone?"

Since the first confirmed discovery of an exoplanet in 1992~\cite{wolszczan1992planetary}, thousands of exoplanets have been found.
The majority of them have been detected using indirect methods, such as the transit method~\cite{ricker2015transiting}, which measures the dimming of a star as a planet passes in front of it, or the radial velocity method~\cite{lovis2010radial}, which detects wobbles in a star’s motion due to gravitational pulls from orbiting planets.
While indirect techniques have been highly successful, they provide limited information about the planets themselves.

Direct imaging of exoplanets is essential for finding planets like our own~\cite{Chauvin2004AGP, Macintosh2014FirstLO}.
With the advent of powerful instruments such as the James Webb Space Telescope (JWST), the Roman Space Telescope, the upcoming class of Extremely Large Telescopes, and the proposed Habitable Worlds Observatory, we are entering a new era of exoplanet exploration, where the discovery and study of Earth-like planets in habitable zones are within reach~\cite{PathwaysAstro2020}.
Compared to indirect methods, direct imaging could allow us to better understand the properties of exoplanets and characterize their atmospheres, as well as provide more details about the system in which they reside.
However, directly imaging exoplanets remains a formidable challenge due to the extreme contrast in brightness between the bright host stars and the faint planets. 
Earth-like planets could be ten billion times fainter than the Sun-like stars they orbit~\cite{Guyon2018ExtremeAO}, making it difficult to distinguish the planet from the overwhelming star glare and the instrumental noise of our telescopes.
Therefore, with current technology, direct imaging is largely constrained to observing planets that are sufficiently far from the star glare or are particularly massive.
Detecting potentially habitable, Earth-like planets remains beyond our current capabilities, and significant technological advancements will be important to push the boundaries of direct imaging toward the detection and characterization of potentially habitable worlds.

Coronagraphy is a key instrumentation technique in high-contrast direct imaging to mitigate the impact of starlight. 
Coronagraphic telescope instruments employ a series of optical elements to suppress the overwhelming starlight while preserving the light from the surrounding planets~\cite{Guyon2018ExtremeAO, Rouan2000Four, Mawet2005AnnularGP}. 
However, even with the advanced coronagraphic systems employed by leading telescopes like JWST, starlight suppression is not perfect, and the residual diffracted starlight produces speckle patterns that further complicate the problem~\cite{Frazin2012UtilizationOT}. 
The speckles can be brighter than the planets of interest and of very similar spatial scales. 
The speckles may also evolve on a variety of timescales depending on their different physical causes, such as defects or alignment drifts in different parts of the telescope instrument. 

In the face of these challenges, post-processing techniques have become a vital tool, in addition to instrumentation advancements.
Current post-processing techniques for exoplanet imaging generally involve three steps: taking reference images, using them to estimate the star's point spread function (PSF), and subtracting this estimated PSF from the science images to reveal the underlying planet signal.
Common methods to collect reference images are Angular Differential Imaging (ADI)~\cite{Marois2005AngularDI}, which observes the same scene at various roll angles, Spectral Differential Imaging (SDI)~\cite{Marois2005TRIDENTAI}, which observes the same scene simultaneously at different wavelengths, and Reference Star Differential Imaging (RDI)~\cite{Lafrenire2007ANA}, which directly observes different stars as reference.
With these reference images, techniques like principal component analysis (PCA) are used to estimate the star PSF, which is then subtracted from the science image~\cite{Soummer2012Detection, Gonzalez2017VIPVI}.

\begin{figure*}[t]
\centering
\includegraphics[width=0.99\linewidth]{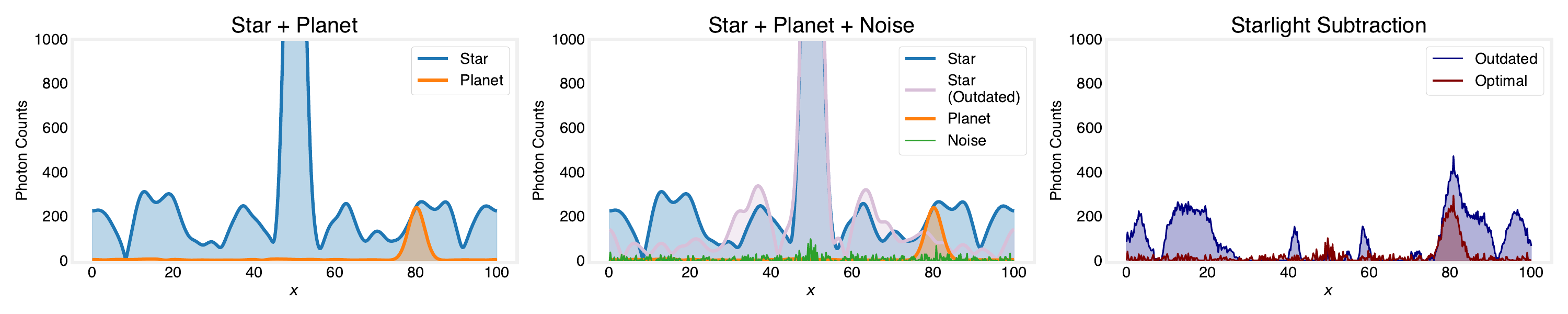}
\caption{A 1D toy example illustrating the basic principles and challenges of the problem addressed in this paper, simplified by examining the PSF in 1D. The first panel shows the photon counts resulting from a star and a planet, including the effects of wavefront aberration. The second panel additionally displays an incorrect star PSF used in practice (measured under a different wavefront aberration), and the level of photon noise in observations. The third panel shows 1) Outdated: subtraction using the outdated star PSF, which leaves significant residuals that can be mistaken for the planet signal, posing the risk of false positive detection or too low signal-to-noise; 2) Optimal: perfect starlight subtraction, representing the upper bound of performance, limited only by noise. The core imaging problem is separating the planet signals from the star signals. 
Our goal is to improve on the outdated starlight subtraction result via differentiable optimization, achieving a more accurate separation of the star and planet signals. This will effectively result in accurately reducing starlight residuals and improving planet detection sensitivity.}
\label{fig:1D_toy}
\end{figure*}

In a nutshell, the post-processing problem for exoplanet imaging boils down to {\it effectively separating signals of the stars from the planets} in the presence of speckle patterns arising from wavefront aberrations, which can mimic planet signals and complicate detection. 
These complications can lead to false positives (when speckles are mistaken for planets) and false negatives due to over-subtraction (when genuine planet signals are removed along with speckles).
Fig.~\ref{fig:1D_toy} provides an intuitive toy example illustrating the underlying challenges of this problem.
While useful, current post-processing techniques have several limitations: they do not leverage wavefront sensing data, which could provide hints about speckle structure; they cannot account for dynamic wavefront aberrations occurring as the telescope moves between observations; they may require additional reference images, consuming valuable telescope time that could be used for scientific observations; and importantly, they do not model the underlying physical phenomena that give rise to the speckle patterns.
The first attempt to address this last limitation \cite{ygouf2013simultaneous} proved computationally challenging due to then-available optimization strategies.

Exoplanet imaging stands to benefit greatly from differentiable rendering techniques, which have recently transformed computer graphics, computer vision, and computational imaging~\cite{Li2018DifferentiablePF, liu2019soft, mildenhall2020nerf}. 
By applying differentiable rendering, we may overcome current post-processing limitations, providing a more comprehensive and adaptable framework for tackling high-contrast exoplanet imaging challenges~\cite{Desdoigts_2023, Desdoigts_2024}.
Differentiable rendering encourages consideration of the entire image formation process rather than focusing solely on final images. 
This perspective naturally leads us to examine the wave space, considering light propagation before it reaches the detector.
{In this paper, we propose a differentiable rendering technique for exoplanet detection, which differs from the conventional approach that focuses on the image intensity domain and relies solely on image detector data.}
We extend exoplanet imaging post-processing considerations to encompass the entire wave space, employing wave-optics modeling for image formation and incorporating wavefront sensing data. 

Our case study focuses on the James Webb Space Telescope (JWST), where wavefront sensing measurements primarily reflect minor misalignments of the telescope's hexagonal mirrors\footnote{These measurements are typically taken every few days, with mirror realignment occurring only when errors exceed a specified threshold. Due to far less frequent corrections compared to image measurements, uncorrected aberrations will still affect detector images between adjustments.}. 
These wavefront sensing data provide valuable aberration estimates for our observations, offering information for a successful star-planet signal separation.
However, the available aberration estimates derived from these wavefront sensing data are generally {\it outdated} for a given image, as numerous science observations can be conducted between two wavefront measurements, which may be separated by several days.
Consequently, by the time a science observation is performed, the underlying wavefront aberration affecting the observed image may have drifted by an unknown amount from the most recent available estimate.

Our differentiable rendering method transforms these outdated aberration estimates into a valuable resource for post-processing, effectively overcoming the temporal mismatch through a gradient-based optimization to refine outdated estimates and improve starlight subtraction.
We construct a differentiable model of the image formation process that takes the estimated wavefront aberration map as input.
This model allows us to compute a reconstruction loss between the rendered image and the captured image, enabling gradients to backpropagate and refine the estimated wavefront aberrations.
By using the refined aberration estimates to render a final star PSF, we obtain a more accurate estimate of the star signals. 
This improved accuracy allows us to surpass the detection limits of traditional post-processing pipelines.

The main contributions of this work are:
\begin{itemize}
    \item We introduce a differentiable renderer with fast and parallelizable GPU implementation for high-contrast imaging, accurately modeling light wave propagation through a coronagraphic optical system.
    \item We leverage differentiability to refine the underlying wavefront aberrations through gradient-based optimization, fully utilizing the wave space information.
    \item We show the effectiveness of our wave space approach through simulation experiments, achieving significant improvements in starlight subtraction and planet detection.
    \item We provide insights into the robustness of our method under different observing conditions and its potential impact on future exoplanet imaging missions.
\end{itemize}

\begin{figure*}
\centerline{\includegraphics[width=0.95\linewidth]{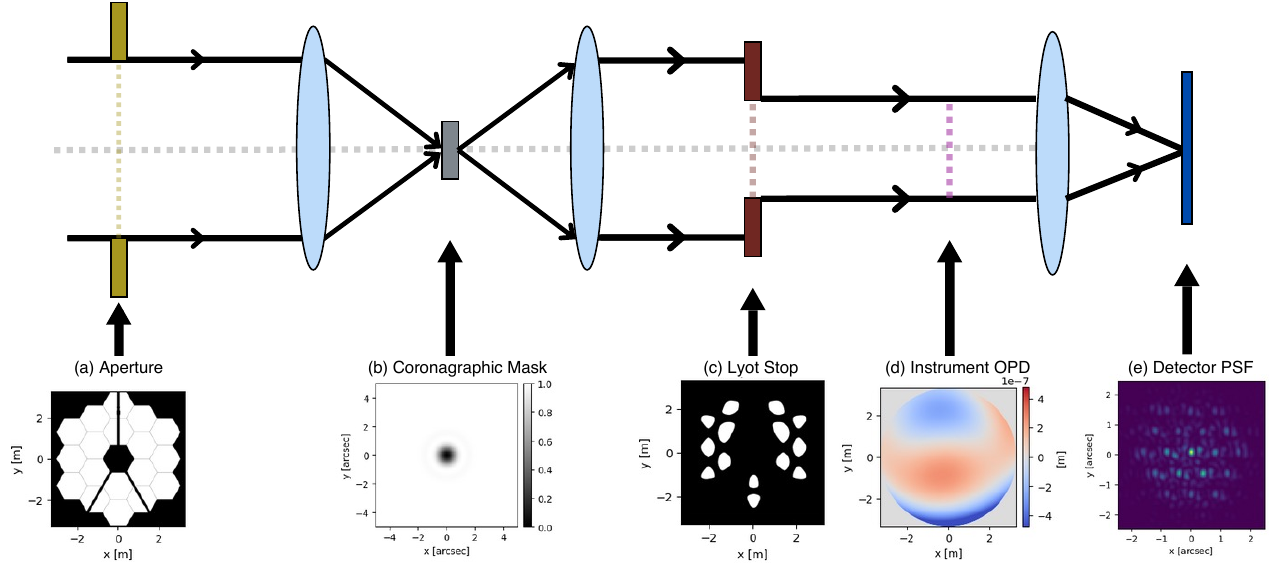}}
\caption{Schematic diagram of the Lyot Coronagraph equivalent to the one in JWST NIRCam, showing the role and impact of each optical element. 
This optical system forms the basis of our differentiable rendering approach and illustrates the complexity of the imaging process we aim to model.
Light enters from the left, reflecting off the primary mirror (aperture) in the pupil plane, then focuses on the image plane where a coronagraphic mask blocks the central region containing the star. 
In the subsequent pupil plane, a Lyot stop suppresses residual diffracted starlight before the light undergoes instrument-induced wavefront aberrations and is refocused onto the image plane detector. 
The plots show (a) aperture amplitude transmission, (b) coronagraphic mask amplitude transmission, (c) Lyot stop amplitude transmission, (d) static optical aberrations in the NIRCam instrument, and (e) resulting point spread function (PSF) at the detector, displaying measured intensity at each pixel.}
\label{fig:coronagraph}
\end{figure*}

\section{Background}

\subsection{Definitions}
To ensure clarity and consistency throughout the paper, we formally define the following key terms:

\underline{Contrast}: 
We use contrast to refer to the brightness ratio between the host star and the orbiting exoplanet. 
It is expressed in a logarithmic scale, with a contrast level of $10^{-x}$ indicating that the planet is $10^x$ times fainter than its host star\footnote{Planets of a contrast level can have different detector intensities depending on their positions in the field of view. 
A planet close to the center of the field of view will be attenuated by the focal plane mask designed to occult the star. 
Thus, the true contrast level of {the planet} is not necessarily reflected by its brightness \textit{on the detector}.
Similarly, the {brightness of the star}, i.e. the peak brightness of the astronomical object unocculted by the coronagraph, is not the peak brightness of the occulted star PSF recorded on the detector.}.

\underline{Aberration}: 
We denote wavefront aberrations as $\phi$, a two-dimensional representation of the phase deviations of a wavefront from an ideal, unaberrated wavefront (i.e., a plane wave). 
This 2D map quantifies the optical path differences across the pupil of the imaging system, with each point in the map corresponding to a specific location in the pupil. 
In real-world systems, wavefront aberration reflects the optical path differences measured by a wavefront sensor and expressed in units of length. 
For JWST, the optical path differences across the primary mirror on the telescope usually have a root mean square value of tens of nanometers.
\red{In this work, during optimization, we represent the wavefront aberrations $\phi$ explicitly as a 2D grid of $1024\times1024$ pixels. Future work may explore parameterizing $\phi$ with Zernike polynomials or other learned or analytical basis functions to potentially reducing the complexity of the optimization.}

\underline{Drift}: 
In this paper, drift refers to the gradual change of the wavefront aberrations over time, which can be caused by factors such as thermal variations, mechanical instabilities, or the movement of optical elements.
For our purpose, it quantifies wavefront aberration from an outdated measurement that has been taken at an earlier timestep.

\underline{SNR (Signal-to-Noise Ratio)}:
In the context of exoplanet imaging, SNR is commonly used to measure the quality of the recovered planet signal,
SNR in exoplanet detection is {\it computed over the annulus of the planet}, rather than the entire image or field of view~\cite{Ruffio2017ImprovingAA}. 
The signal term refers to the planet brightness in the final subtracted image, and the noise term is the standard deviation of pixel values in the planet's annulus (width approximating the spatial size of a planet), excluding the planet signal.
We adopt this annulus SNR to quantitatively assess the planet signal recovery in accordance with existing exoplanet imaging literature~\cite{Ruffio2017ImprovingAA}, where the standard for a valid planet detection is to have an annulus SNR above 5.

\subsection{Coronagraphy}
\label{sec:background_corona}
Coronagraphy is a key technique employed in high-contrast imaging to suppress the overwhelming starlight and enhance the detectability of faint exoplanets. 
Fig.~\ref{fig:coronagraph} illustrates a schematic of a coronagraphic system used in the JWST NIRCam instrument. 
Coronagraphic instruments typically consist of a series of optical elements, including apodizers, focal plane masks, and Lyot stops.
These optical elements are designed to selectively block and filter the light from the central star while preserving the light from the surrounding planets~\cite{Guyon2018ExtremeAO, Rouan2000Four, Mawet2005AnnularGP}.

As shown in Fig.~\ref{fig:coronagraph}, after the entrance pupil (a $6.5$ meter mirror composed of smaller individual hexagonal mirrors), a focal plane mask blocks the central starlight in the image plane. 
Then, a Lyot stop filters out the remaining diffracted light in the pupil plane before the final science image is formed on the detector.
The performance of coronagraphic systems is critically dependent on the wavefront quality of the incoming light. 
Wavefront aberrations, caused by factors like atmospheric turbulence, optical defects, and mechanical disturbances, produce speckles in science images that can resemble exoplanets~\cite{Beuzit2019SPHERETE, Frazin2012UtilizationOT}.

\subsection{Wavefront Aberration Sources}
\begin{figure*}[t]
\centering
\includegraphics[width=0.99\linewidth]{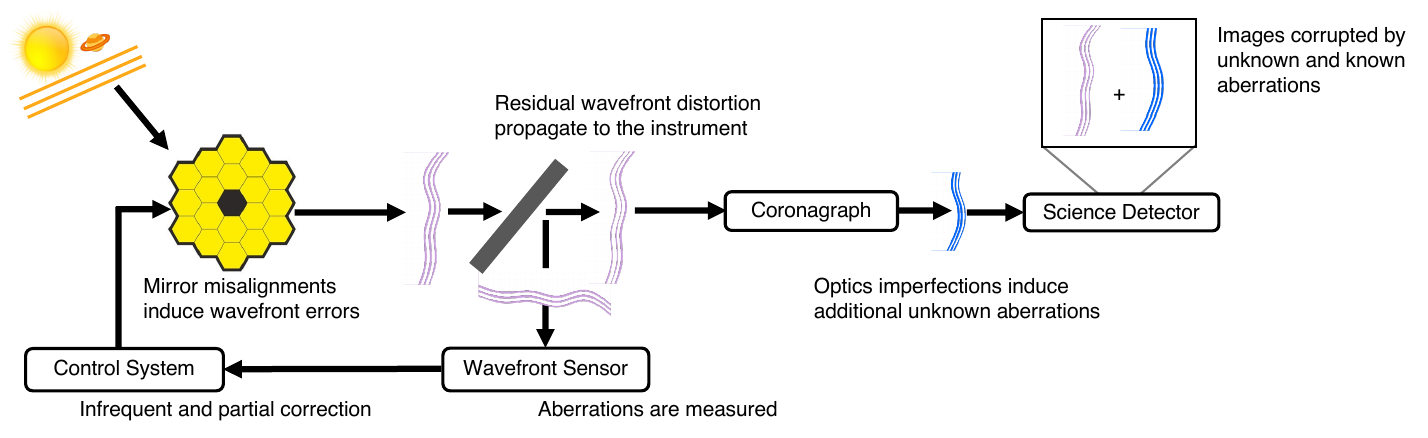}
\caption{Illustration of the sources of wavefront aberrations. In space-based instruments like the James Webb Space Telescope (JWST), aberrations stem from mirror misalignments, deformations, or tilt errors (purple curves) and slowly-evolving non-common path aberrations (blue curves);
the former are measured typically every two days (not simultaneous with the science observations), while the latter are not measured directly. 
See the Appendix \ref{s:appendix_A} for details specific to JWST and NIRCam. Wavefronts are partially corrected by deformable mirrors, but residual distortions persist in the final image due to optical imperfections.}
\label{fig:scenarios}
\end{figure*}

While ground-based instruments primarily contend with rapidly evolving atmospheric turbulence, space-based telescopes like JWST face a different set of issues. 
As shown in Fig.~\ref{fig:scenarios}, aberrations in space primarily stem from factors such as mirror misalignments, thermal distortions, and mechanical vibrations, evolving over longer timescales ranging from minutes to days.
A critical source of wavefront aberrations common to both ground-based and space-based instruments is non-common path aberrations (NCPAs)~\cite{Sauvage2010SAXOTE}. 
These aberrations occur in the optical path unique to the science camera downstream of the wavefront sensor. 
Consequently, NCPAs are neither directly measured by the wavefront sensor nor perfectly corrected by adaptive optics systems.
The result is the presence of static or slowly evolving speckles in science images, which can mimic or obscure potential exoplanet signals.

This paper focuses specifically on space-based scenarios, where the potential for long, uninterrupted observations without atmospheric turbulence provides unique opportunities for discovering and characterizing new exoplanets. 
The wavefront aberration sources considered in this work are primarily those associated with telescope optics, which are the dominant factors limiting the performance of space-based high-contrast imaging systems.
Specifically, three types of aberrations are included: known but outdated primary mirror aberrations, known static instrumental aberrations, and unknown NCPAs.
We take the JWST as our case study due to 1) its unprecedented sensitivity to new exoplanet populations, 2) its extraordinary stability, and 3) the wide availability of wavefront data for the telescope from both in-lab testing and in-flight measurements.

\begin{figure*}[t]
\centerline{\includegraphics[width=0.98\linewidth]{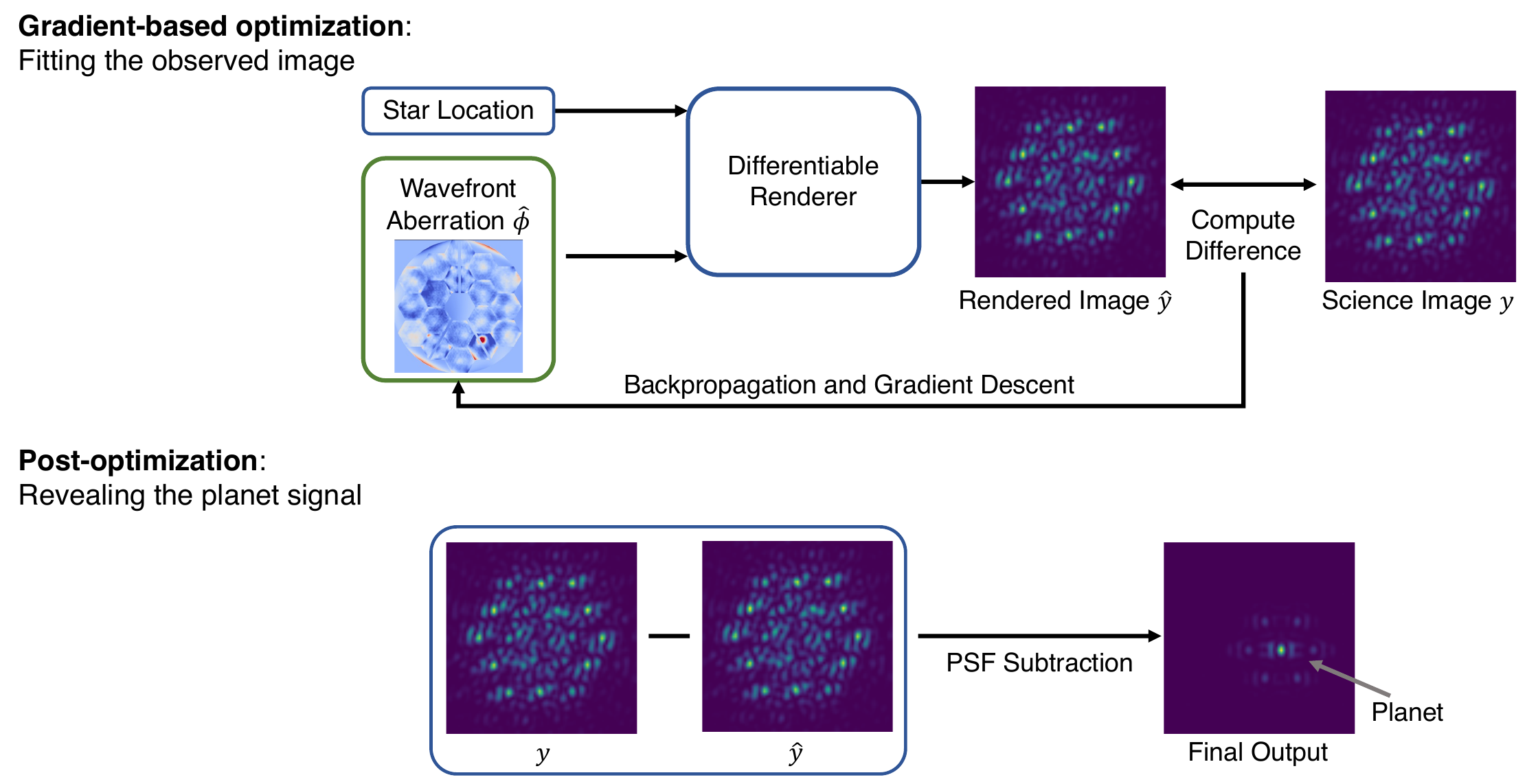}}
\caption{Overview of the proposed differentiable rendering framework for enhanced exoplanet imaging. 
The method employs gradient-based optimization to improve the initial, outdated wavefront aberration estimate $\hat{\phi}$ based on the discrepancy between the rendered image $\hat{y}$ and the observed science image $y$. 
The differentiable renderer facilitates efficient backpropagation and gradient descent updates. 
It is important to note that while $y$ contains signals from both the star and the planet (along with noise and background), $\hat{y}$ only considers the star PSF. 
Since the optimization changes the structure of the estimated wavefront aberration, which primarily affects the star PSF, it is unlikely to overfit localized features like a planet as it would introduce larger inconsistencies with the dominant star signals.
Upon convergence, the refined star PSF estimate is used for PSF subtraction, producing a final image where the planet signals are significantly enhanced and speckle noise is effectively suppressed.}
\label{fig:pipeline_overview}
\end{figure*}

\subsection{Post-Processing Techniques}
Various post-processing techniques have been developed to estimate and then subtract the PSF from the science images, revealing the planet signal while mitigating the impact of wavefront aberrations and speckle noise.

One commonly used approach is Angular Differential Imaging (ADI)~\cite{Marois2005AngularDI}, which exploits the field rotation in altitude-azimuth telescopes to distinguish between the static speckles and the rotating planet signal.
By taking a sequence of images at different sky angles and combining them using techniques such as Locally Optimized Combination of Images (LOCI)~\cite{Lafrenire2007ANA} or Karhunen-Loève Image Projection (KLIP)~\cite{Soummer2012Detection}, the star PSF can be estimated and subtracted.

Spectral Differential Imaging (SDI)~\cite{Marois2000Efficient}, leverages the wavelength dependence of speckles to separate them from the planet signal. 
By imaging the target at multiple wavelengths, SDI can exploit the fact that speckles scale with wavelength while the planet signal remains relatively constant, allowing for effective speckle suppression.

Reference Star Differential Imaging (RDI)~\cite{Lafrenire2007ANA} is useful for space-based instruments, where the stability of the PSF is higher. 
RDI relies on multiple observations of reference stars to build a library of reference PSFs, which can then be used to model and subtract the star PSF from the target images.

While these post-processing techniques have been successful in detecting exoplanets, current estimation of the star PSF demands devoting valuable in-flight time and resources to collect reference images, which by themselves are typically not of scientific interest.
These techniques often fail to reach the fundamental contrast limit of detection, leaving potentially detectable planets undiscovered. 
As we show later in Fig.~\ref{fig:noiselimits}, our method approaches the fundamental contrast limit more closely, offering improved sensitivity for exoplanet detection.

\subsection{Differentiable Rendering}

Differentiable rendering of visual data has transformed various domains, including computer graphics and computational imaging, by enabling the optimization of parameters through gradient descent.
This paradigm has unlocked previously intractable inverse problems in imaging and enabled efficient computations to obtain solutions.

In computer graphics, differentiable programming has enabled the rendering of photorealistic images from parametric scene descriptions via iterative optimizations of the rendering parameters to match target images~\cite{Li2018DifferentiablePF, Li2018DifferentiableMC, liu2019soft, mildenhall2020nerf, kerbl20233d}.
Similarly, in computational imaging, differentiable simulation has revolutionized the design of optical systems by enabling the optimization of optical system parameters to achieve desired imaging characteristics~\cite{ Robidoux2021EndtoendHD, Peng2019LearnedLF, Sun2021EndtoendCL}.
Differentiable simulation also facilitates research on new data structures for representing images or videos in imaging and graphics applications that enable faster optimization convergence or more flexible data priors~\cite{Sitzmann2020ImplicitNR, muller2022instant, Mao2021AcceleratingAT, Liu2021RecoveryOC, Feng2023NeuWSNW, Zhou2023FPMINRFP}.
Our work builds upon these advancements by introducing a differentiable renderer specifically designed for the challenges of high-contrast astronomical imaging, which is crucial for the detection of exoplanets.

\red{
\subsection{Connection to Aberration Estimation in Microscopy} 
Our approach to using wavefront aberration data in exoplanet imaging shares similarities with methods in microscopy that leverage guide stars or reference light sources to correct optical aberrations in thick tissues. 
In microscopy, adaptive optics techniques~\cite{Horstmeyer2015GuidestarassistedWM, Vellekoop2015FeedbackbasedWS, Feng2023NeuWSNW} often use these reference points to estimate and correct aberrations, thereby enhancing the clarity of structures in complex, scattering environments.
Similarly, our differentiable rendering framework utilizes wavefront sensing data to mitigate aberrations and improve planet signal recovery, though adapted to the unique challenges of high-contrast space-based imaging.
}

\section{Method}
\label{sec:our_method}

In this section, we present the overall framework of our approach to the problem and provide details of our differentiable renderer, which is designed to model the propagation of light through a coronagraphic optical system.

\subsection{Enhancing Starlight Subtraction}

Traditional post-processing techniques aim to remove starlight based on reference images and reveal the faint exoplanet signal. 
However, these methods overlook a valuable resource: available wavefront sensing data. 
This data directly relates to speckle formation, which often mimics and obscures planet signals. 
In challenging scenarios where the planet signal is weak relative to the speckle noise, the information offered by wavefront data could be important for planet detection. Not utilizing this already-collected information represents a missed opportunity to maximize observational data.

\begin{figure*}[ht]\centerline{\includegraphics[width=0.95\linewidth]{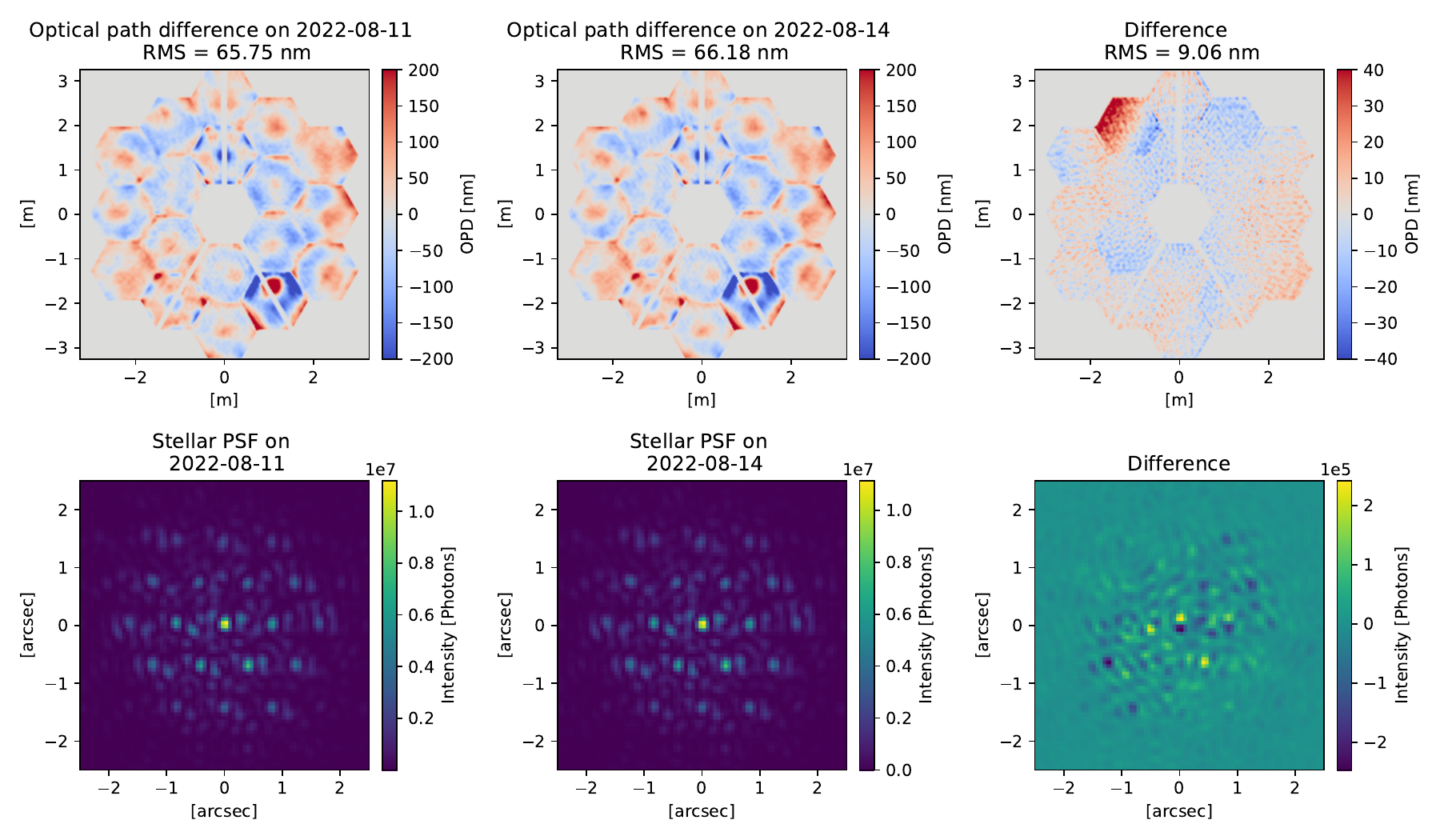}}
\caption{Visualization of wavefront aberration evolution with real JWST data.
We show consecutive optical path difference measurements (taken around three days apart) and their resulting star PSFs, as well as the difference between both. 
The figure displays consecutive optical path difference (OPD) measurements taken approximately three days apart (a)-(b), their resulting star point spread functions (PSFs) (d)-(e), and the differences between both measurements (c),(f).
The OPD color scale is limited to $(-200, 200)$ nm to highlight the overall structure, although a significant meteoroid impact on the lower right mirror segment caused local OPD values to reach 800 nm.
This visualization demonstrates the temporal evolution of wavefront aberrations and their impact on the star PSF.}
\label{fig:opds_psfs}
\end{figure*}

To address these challenges and advance exoplanet imaging and detection, {we propose a post-processing approach based on differentiable rendering}.
By constructing a differentiable model of the image formation process of a coronagraphic imaging system, we enable the inversion of the optical forward model.
This allows us to refine outdated wavefront aberration estimates, extracting untapped value from the wavefront space.

The proposed framework (Fig.~\ref{fig:pipeline_overview}) establishes a direct and differentiable connection between the wavefront aberration and the observed science images.
A key insight of our approach is its focus on accurately estimating the star PSF rather than explicitly modeling the planet signal.
By addressing the overwhelming starlight, which is the dominant source of interference in planet detection, accurate star PSF subtraction naturally reveals any planet signals present without explicitly modeling them.
This method is inherently resistant to overfitting planet signals: attempts to create planet-like features by manipulating wavefront aberrations would introduce larger inconsistencies in the dominant star signal, resulting in a higher overall fitting error.
Moreover, this approach can handle multiple planets or extended structures without modification, as it makes no assumptions about the number of planets.

\subsection{Details of Differentiable Renderer}

Our differentiable renderer, $\mathcal{G}$, models the complex interactions of light within a high-contrast imaging system. 
It takes the central star's position $p_{\text{star}} \in \mathbb{R}^2$ and a wavefront aberration map $\phi \in \mathbb{R}^{H_{\phi} \times W_{\phi}}$ as inputs, rendering a science image $\hat{y} \in \mathbb{R}^{H_{y} \times W_{y}}$ that closely simulates the actual measurements:
\begin{equation}
\hat{y} = \mathcal{G}(\phi,p_{\text{star}}) .
\end{equation}
In practice, we start with an outdated estimate of the wavefront aberrations, $\hat{\phi}$, obtained from the most recent wavefront sensing data. Our goal is to refine this estimate $\hat{y} = \mathcal{G}(\hat{\phi}, p_{\text{star}})$ to better match the observed focal plane image $y$.

The forward model of $\mathcal{G}$ can be expressed as a series of operations:
\begin{equation}
\begin{aligned}
\hat{y}^{noiseless} &= |\mathcal{F}{D \cdot C \cdot \mathcal{F} B \cdot \mathcal{F}{P \cdot (x^{(0)} \cdot e^{i p_{\text{star}}} \cdot e^{i \hat{\phi}})}}|^2 \\
\hat{y} &= \hat{y}^{noiseless} + \epsilon_{shot} + \epsilon_{read}
\end{aligned}
\end{equation}
where $x^{(0)}$ is the input complex wavefront (typically a plane wave), $e^{i p_{\text{star}}}$ applies a phase shift for the star position, $e^{i \hat{\phi}}$ incorporates the estimated wavefront aberrations, $\mathcal{F}$ denotes the Fourier transform, $P$ is the entrance pupil function, $B$ is the focal plane mask, $C$ is the Lyot stop function, $D$ is the NIRCam instrument OPD, $\epsilon_{shot} \sim \text{Poisson}(\lambda = \hat{y}^{noiseless})$
and $\epsilon_{read} \sim \mathcal{N}(0, \sigma_{\text{read}}^2)$.

The optimization objective is to minimize the difference between the rendered and observed images, given the current estimate of the wavefront aberrations.
Formally, we seek to solve the following optimization problem:
\begin{equation}
\hat{\phi} = \operatorname*{argmin}_{\phi} \mathcal{L}(\hat{y}_{ | \phi, p_{\text{star}}}, y),
\end{equation}
where $\mathcal{L}$ is a suitable distance metric. We adopt the $L_1$ norm in our experiments.

By leveraging the differentiability of $\mathcal{G}$, we compute the gradients of the loss function with respect to $\hat{\phi}$ using automatic differentiation: $\nabla_{\hat{\phi}} \mathcal{L} = \frac{\partial \mathcal{L}}{\partial \hat{y}} \frac{\partial \hat{y}}{\partial \hat{\phi}}.$
We then perform iterative gradient descent updates to refine the estimate of the wavefront aberrations:$
\hat{\phi} \leftarrow \hat{\phi} - \eta \nabla_{\hat{\phi}} \mathcal{L}$,
where $\eta$ is the learning rate.

\begin{figure}[t]
\centerline{\includegraphics[width=0.95\linewidth]{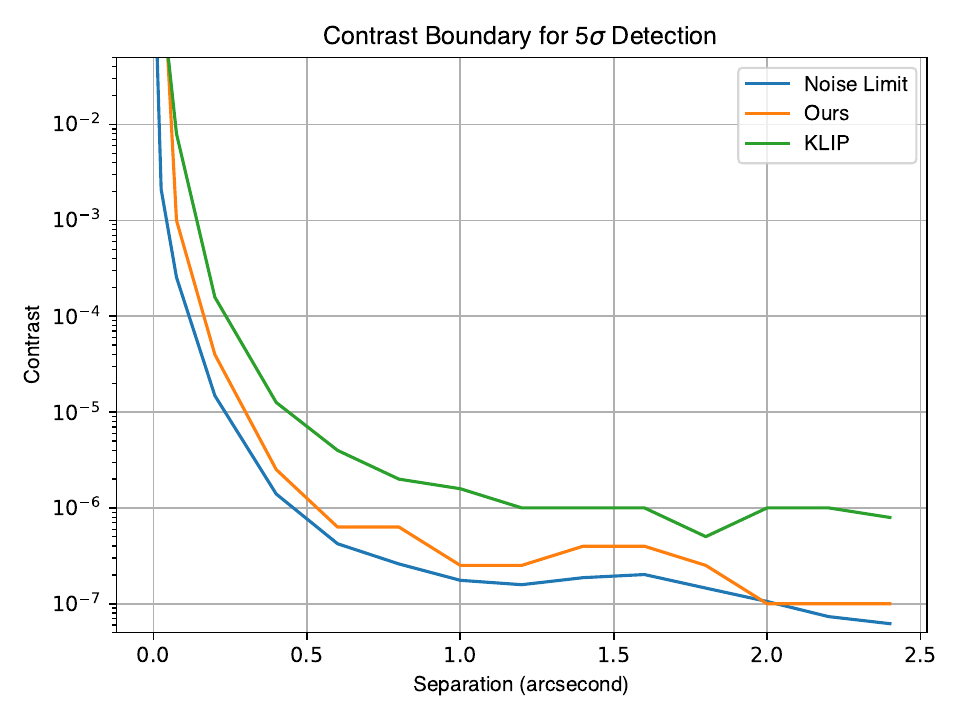}}
\caption{Contrast boundary for $5\sigma$ detection as a function of angular separation from the star.
For a point on a curve, any planet with a contrast (y-axis) below it would have an SNR less than five and thus be considered undetectable at that separation.
The blue curve represents the fundamental noise limit, determined by shot noise and read noise, which sets the theoretical boundary for planet detection. These are the noise limits for the experimental setup assumed in this paper (i.e., the star brightness and spectral type, total observation time, and the filter used).
This noise limit curve is inferred based on the overall noise amount, giving the minimum signal strength required to exceed the noise by a factor of 5.
The green curve shows the performance of KLIP~\cite{Soummer2012Detection}.
The orange curve shows the empirical contrast levels achieved by our method, obtained through iterative optimization at each separation in simulation until the SNR drops below 5. 
Our approach closely approaches the fundamental noise limit across a range of angular separations, demonstrating its effectiveness in maximizing planet detection capabilities within theoretical constraints. 
The proximity of the blue and the orange curves illustrates how our framework nearly achieves the optimal performance allowed by the noise limits. 
%The rise in the contrast limit in the KLIP curve at $\sim$ 2 arcseconds is due to artifacts in the image subtraction from using reference images near the size of the stellar PSF and with a relative shift from one another (see Appendix \ref{s:appendix_C}).
}
\label{fig:noiselimits}
\end{figure}

\begin{figure*}[t]
\centerline{\includegraphics[width=0.98\linewidth]{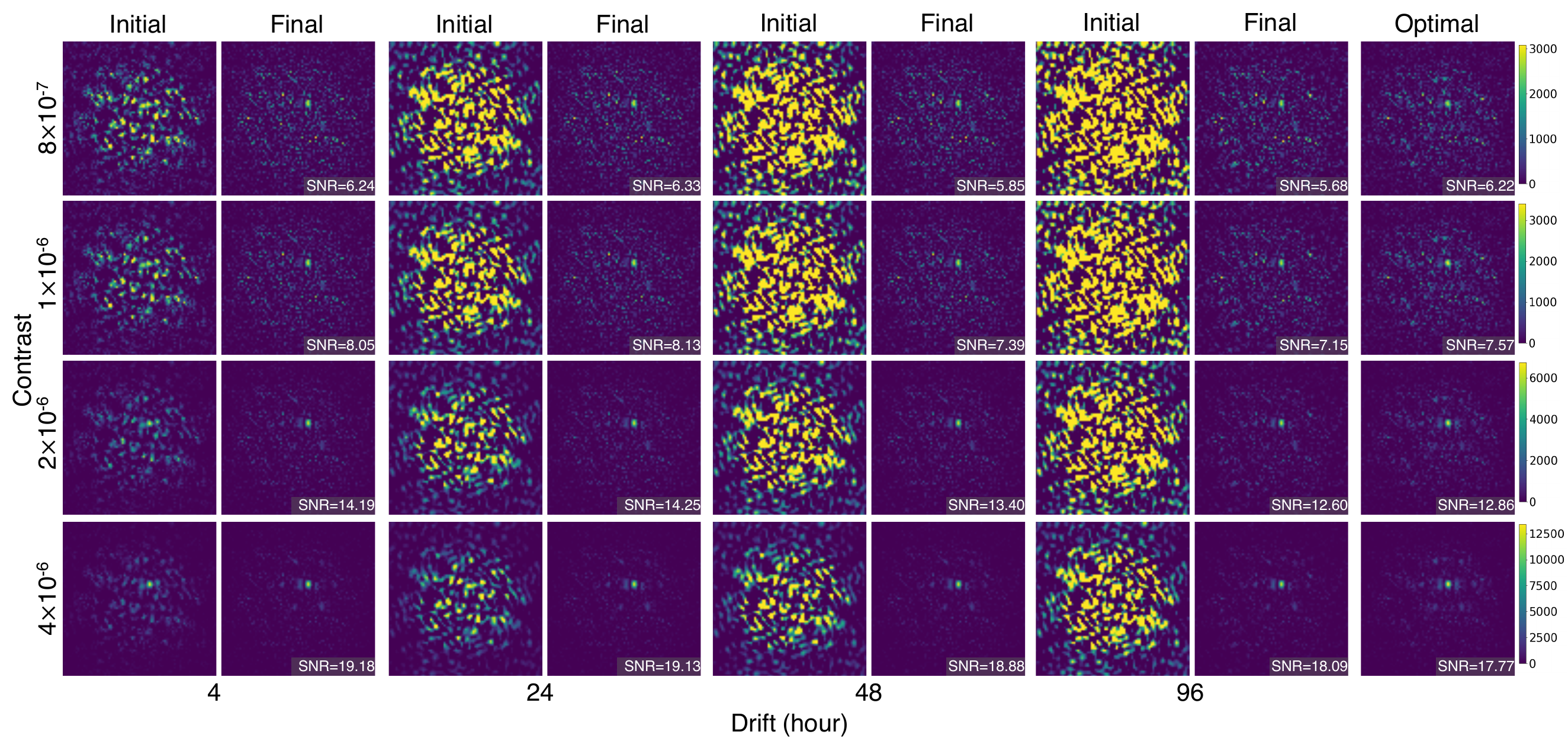}}
\caption{
Effectiveness in improving starlight subtraction at various star-planet contrast and wavefront drift levels.
The angular separation between the planet and the star is fixed at 0.6 arcseconds.
The overall field of view diameter is 5 arcseconds (80 pixels).
Each row shows different contrast levels between the star and the planet ($8\times10^{-7}$ to $4\times10^{-6}$), while each column represents increasing levels of drift (higher drift indicates a larger discrepancy between the previous wavefront sensor measurement and the actual wavefront aberration).
The {\it Initial} images depict the results of subtracting an inaccurate star PSF derived from the outdated wavefront measurement, while the {\it Final} images show the improved starlight subtraction achieved through the optimization process. The {\it Optimal} images illustrate the ideal outcome of subtracting the ground truth star PSF, with imperfections due to measurement noise. 
Note that our method sometimes achieves a higher SNR than the {\it Optimal} result due to over-subtraction of noise, incorrectly attributing some noise to the star PSF. While this results in numerically higher SNR, it does not necessarily indicate a more accurate starlight subtraction and planet recovery.
The differentiable framework consistently improves the contrast and detectability of the planet signal across a wide range of scenarios.}
\label{fig:contrast_v_drift}
\end{figure*}

\begin{figure*}[ht]
\centerline{\includegraphics[width=0.99\linewidth]{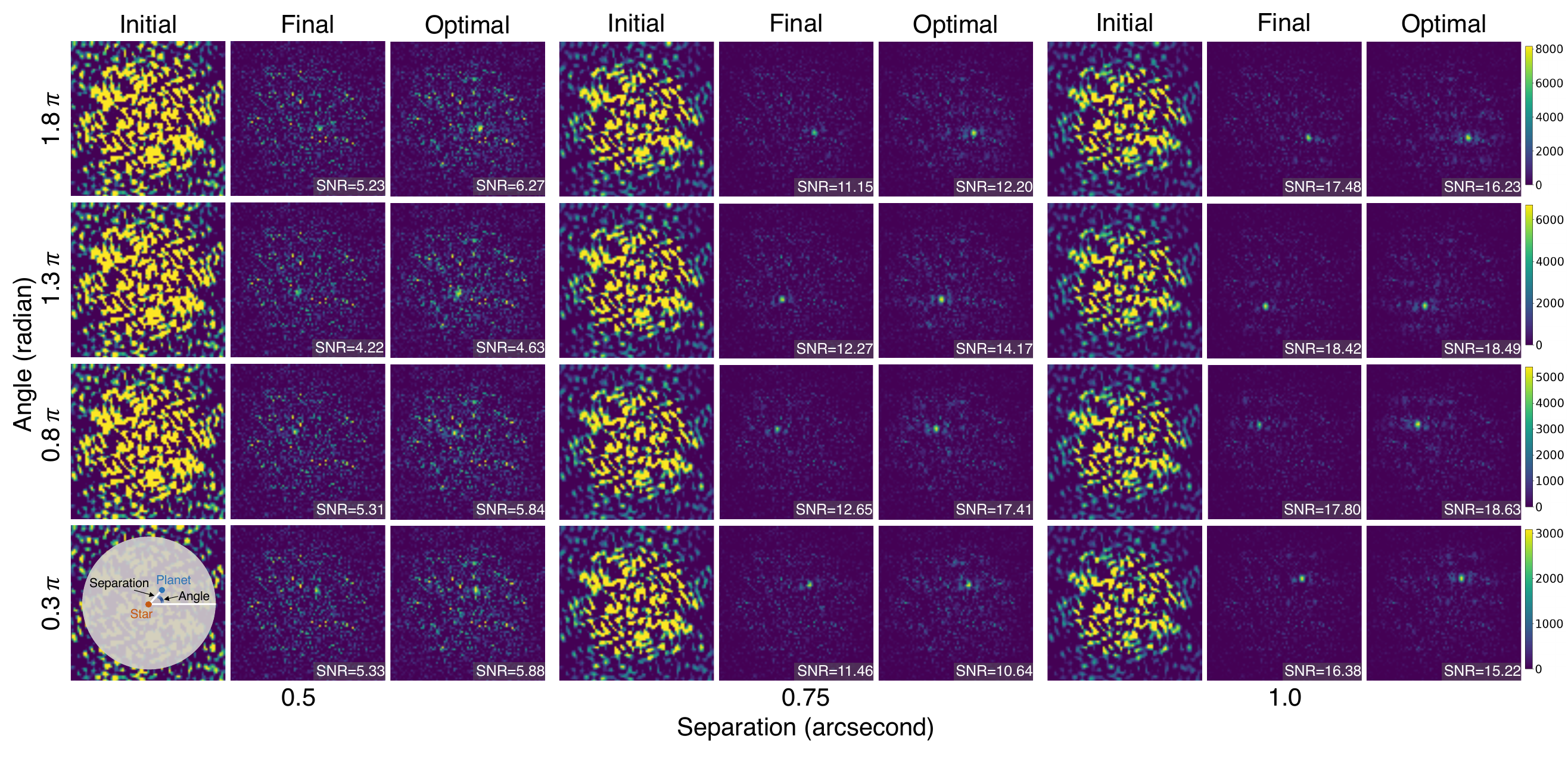}}
\caption{Robustness to variations in the planet's location. 
Separations are measured from the center of the image, which in these experiments is equivalent to the star's position.
The overall field of view diameter is 5 arcseconds (80 pixels).
For a fixed contrast level ($10^{-6}$) and wavefront aberration drift (40 hours), the planet is placed at different locations.
The \textit{Initial} images show the results of subtracting an inaccurate star PSF derived from the outdated wavefront measurement. 
The \textit{Final} images display the improved starlight subtraction achieved through our optimization process. 
The \textit{Optimal} images illustrate the ideal outcome of subtracting the ground truth star PSF, with imperfections due to measurement noise.
Note that our method sometimes achieves a higher SNR than the {\it Optimal} result due to over-subtraction of noise, incorrectly attributing some noise to the star PSF. While this results in numerically higher SNR, it does not necessarily indicate a more accurate starlight subtraction and planet recovery.
The results show that our method consistently achieves effective starlight subtraction and preserves the planet signal across all tested locations, highlighting its robustness to uncertainties in the planet's position.}
\label{fig:varying_location}
\end{figure*}

\begin{figure*}[t]
\centerline{\includegraphics[width=0.99\linewidth]{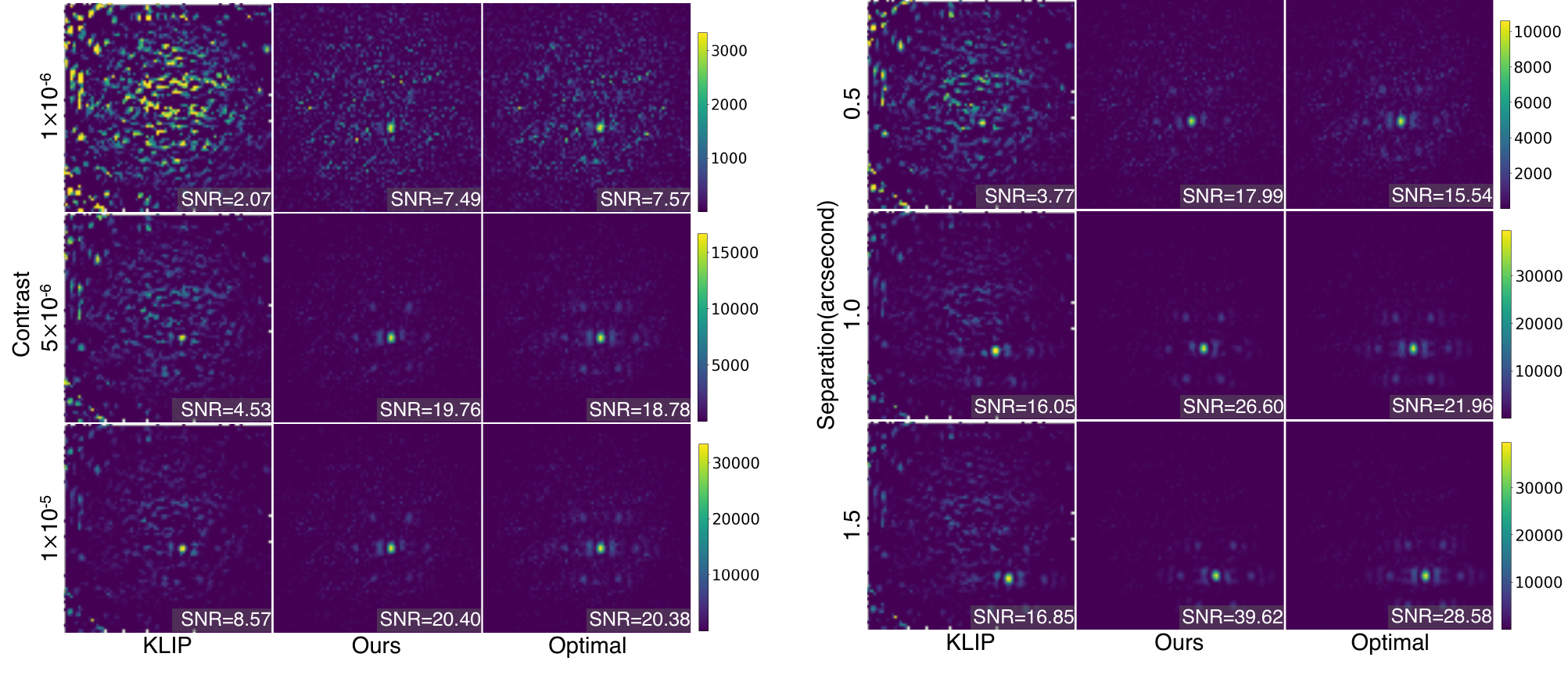}}
\caption{Comparison of our method against KLIP. Left: Performance across various star-planet contrast levels ($1\times10^{-6}$ to $1\times10^{-5}$) at a fixed separation of 0.6 arcseconds and 40 hours of wavefront aberration drift. Right: Performance across different angular separations (0.5 to 1.5 arcseconds) at a fixed contrast of $5\times10^{-6}$ and 40 hours of drift. The {\it Optimal} column shows the subtraction result using the ground truth star PSF. Our method consistently outperforms KLIP, particularly at higher contrast ratios and smaller separations, demonstrating improved planet signal recovery in challenging scenarios. Note that our method sometimes achieves a higher SNR than the {\it Optimal} result due to over-subtraction of noise, incorrectly attributing some noise to the star PSF. While this results in numerically higher SNR, it does not necessarily indicate a more accurate starlight subtraction and planet recovery.}
\label{fig:klip_comparison}
\end{figure*}
\begin{figure*}[!t]
\centerline{\includegraphics[width=0.95\linewidth]{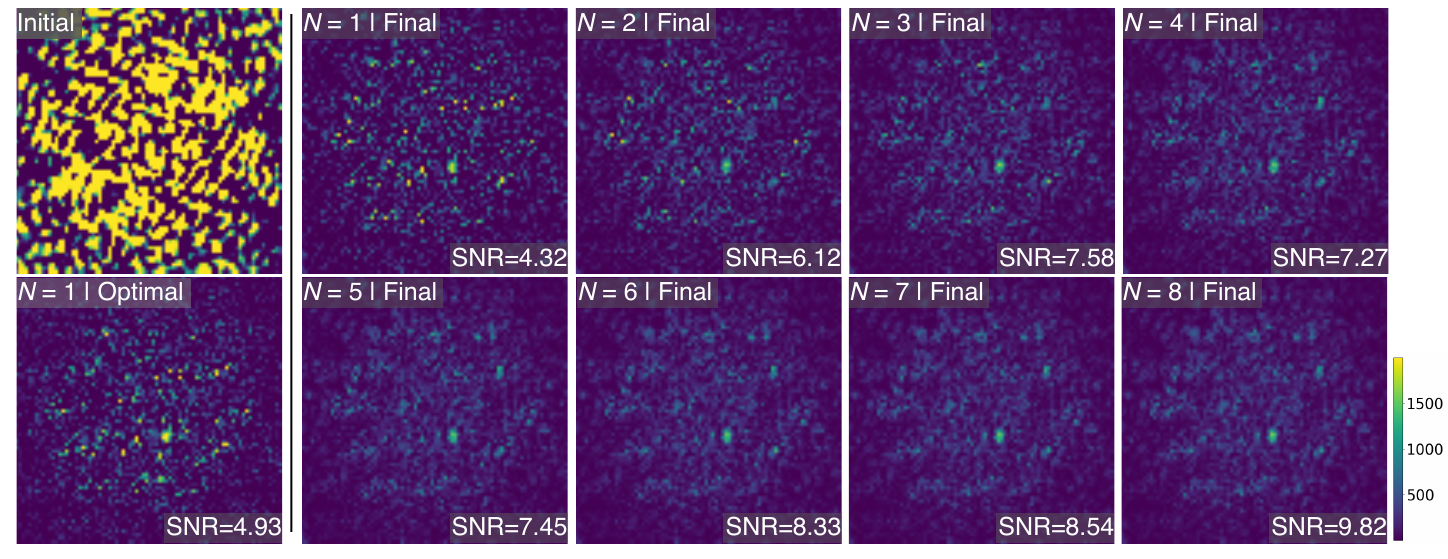}}
\caption{Performance with multiple image measurements of the same planet-star configuration. 
Each image measurement is taken under a different wavefront aberration, generated by linearly interpolating between two real JWST wavefront aberration measurements with different, linearly increasing drift values.
We separately solve for the aberration estimate for each measurement and display the averaged subtraction result here.
The top left panel shows the initial subtraction (0.6 arcseconds separation, contrast $6\times10^{-7}$, 80 hours of wavefront aberration drift). 
The bottom left panel shows the {\it Optimal} subtraction (with noise still present), assuming the ground truth star PSF is available.
While the subtraction for the single-measurement case ($N = 1$) is the primary focus of this paper, the subsequent panels ($N = 2$ to 8) display results for scenarios where multiple image measurements are available.
As the number of measurements increases, the optimization process can better overcome the noise and resolve the exoplanet signal, outperforming the $N=1$ case. These results demonstrate the potential for improved exoplanet detection capability when multiple image measurements can be considered.}
\label{fig:multi_snapshot}
\end{figure*}

Our approach integrates differentiable programming with physics-based modeling in the wave space. 
Going beyond classical image-space approaches, our wave space perspective allows us to construct more accurate models of the star PSF, better capture the effects of subtle aberrations, and more successfully separate them from faint planet signals.

\section{Simulation Experiments}
To validate the potential of our proposed approach for enhanced exoplanet imaging, we conducted a series of simulation experiments. 
The primary objectives were to assess its effectiveness in improving starlight subtraction and to evaluate the robustness of the method under variations in the planet location and contrast level.

\subsection{Setup}
We aim to replicate challenging high-contrast imaging scenarios under varying levels of wavefront aberration. 
For each simulated scenario, we place the star behind the coronagraph and a planet at a specified location $p_{planet}$ and at a chosen contrast level. 
The wavefront from each point source is propagated through the coronagraph optics, including the wavefront aberrations. The designs of the coronagraph optics and processing for the wavefront aberrations were based on \texttt{webbPSF}~\cite{Perrin_2014}, a PSF simulator dedicated to JWST instruments maintained by the Space Telescope Science Institute, which leads the science and mission operations of JWST.
See Appendix \ref{s:appendix_A} for details on simulation and calibration.

For simplicity, the simulations were carried out assuming monochromatic point sources with a wavelength of $4.5$ $\mu$m. 
This wavelength is approximately the center of the F444W filter used in JWST's NIRCam instrument, which is the primary filter used for exoplanet direct imaging. 

To emulate the evolution of wavefront aberrations, we used two consecutive (86 hours apart) JWST in-flight optical path difference (OPD) measurements, one from August 11, 2022, and one from August 14, 2022 (Fig.~\ref{fig:opds_psfs} shows the two real measurements, the magnitude of their differences, and the resulting star PSFs). 
We processed them in the default manner of \texttt{webbPSF} (see Appendix \ref{s:appendix_A}) to obtain two OPD maps for the aperture of the telescope. 
The first one was used as the measured or ``outdated'' OPD map $\hat{\phi}$.
The ``ground truth'' aberration $\phi$  was obtained by interpolating between the outdated OPD map $\hat{\phi}$ and the second real JWST wavefront aberration measurement, with the linear interpolation weight setting the ``drift'' of the actual aberration away from our outdated wavefront measurement sampled at an earlier timestep. 

We consider two sources of image noise: shot noise and read noise.
Shot noise is the primary source of noise, which follows a Poisson distribution with rate $\lambda = \sqrt{n_{\text{photons}}}$ (where $n_{\text{photons}}$ is the number of photons at each pixel).
An additional source of noise is detector read noise, which is much less significant than shot noise.
We model read noise as independent Gaussian noise at each pixel (see Appendix \ref{s:appendix_A}).
These noise levels were calibrated using realistic assumptions of observation strategies and astronomical objects of interest, as well as JWST and NIRCam characteristics (see Appendix \ref{s:appendix_A}). 

The optimization process used the Adam optimizer implemented in PyTorch~\cite{paszke2019pytorch}, with a learning rate of $\eta=10^{-11}$ under a mean absolute error loss function, and ran for 1,000 iterations for each simulated case, which amounted to less than a minute of runtime on an NVIDIA V100 GPU.
The low learning rate reflected the scale of our learnable parameters, the wavefront aberrations, which represent the optical path difference of light with a micrometer-scale magnitude.
It also served as a form of regularization, preventing the optimization from straying too far from the initial aberration estimate, which is based on the most recent wavefront sensing data.

\subsection{Results}
Fig.~\ref{fig:noiselimits} illustrates the detection capabilities of our method in comparison to the fundamental noise limits as a function of angular separation from the star. 
The blue curve represents the theoretical detection boundary set by fundamental noise sources.
The orange curve shows the empirical performance of our method, determined through iterative simulations at each angular separation.
The green curve shows the empirical performance of the state-of-the-art method KLIP~\cite{Soummer2012Detection}.
Our approach outperforms KLIP and nearly achieves the optimal performance allowed by fundamental noise limits across a wide range of angular separations.

Fig.~\ref{fig:contrast_v_drift} demonstrates the effectiveness of our approach in refining the star PSF estimate for improved starlight subtraction. 
Each row corresponds to a different contrast level between the star and planet at a fixed separation of 0.6 arcseconds, and each column represents varying levels of drift (higher drift indicates a larger discrepancy between the outdated wavefront sensor measurement and the actual wavefront aberration).
In each cell, the ``Initial'' images depict the subtraction results using an inaccurate star PSF derived from the latest wavefront measurement,
while the ``Final'' image shows the improved starlight subtraction and enhanced planet signal detection achieved after the optimization process. 
The refined starlight subtraction effectively removes the speckles while preserving the planet signal, as confirmed by the ``Optimal" image, which shows the best-case scenario where we perform starlight subtraction using the ground truth star PSF.
These results showcase the ability of our approach to significantly enhance the detectability of faint exoplanet signals.

Fig.~\ref{fig:varying_location} demonstrates the robustness of our approach to variations in the planet location within the image. 
For a fixed contrast level ($1\times10^{-5}$) and wavefront aberration drift of 40 hours, we placed the planet at different locations within the image. 
The results show that our method consistently achieves effective starlight subtraction and preserves the planet signal across all tested locations. 
The performance does degrade as the planet's position approaches the central star, as expected in high-contrast imaging. 
This degradation is due to the increasing dominance of the star PSF and speckle noise closer to the center, making planet detection more challenging.
\red{
In Fig.~\ref{fig:contrast_v_drift} and Fig.~\ref{fig:varying_location}, our method occasionally achieves higher SNR values than the ``Optimal" image due to over-subtraction of noise, where some noise is incorrectly attributed to the star PSF. 
However, this does not necessarily indicate a more accurate starlight subtraction or improved planet recovery.
}

Fig.~\ref{fig:klip_comparison} shows comparisons between our method and the widely used post-processing method KLIP~\cite{Soummer2012Detection}. 
Since KLIP requires a set of reference images, we replicated an RDI scenario by simulating an appropriate reference star at 9 slightly different sub-pixel positions.
This choice closely follows the 9-POINT-CIRCLE dithering strategy used in real observations~\cite{2016SPIESmallGridDithers,NIRCamSmallGridDithers}, thus obtaining 9 reference images (see Appendix \ref{s:appendix_C} for details). 
To mimic realistic observing conditions, we introduced a time difference of only 3 hours between the reference star and science star observations in terms of wavefront aberration drift. 
This time difference replicates a typical scenario where reference star observations are taken close in time to the science target to minimize variations in observing conditions.
KLIP was performed using the pyKLIP package~\cite{Wang_2015}, using 9 Karhunen-Loeve bases, dividing the image into 5 annuli, and further dividing into 4 subsections in each annulus. 
These hyperparameters were chosen to optimize the PSF subtraction.
We tested both methods under varying contrast levels ($1\times10^{-6}$ to $1\times10^{-5}$) and angular distances between planet and star (0.5 to 1.5 arcseconds), with a wavefront aberration drift of approximately 40 hours.
Leveraging wavefront sensing data and performing targeted optimization of the wavefront aberration estimates, our method demonstrates superior performance across all tested scenarios compared to KLIP, which relies solely on the statistics of the observed science images. 
%Future work improving renderer (e.g., rendering broadband images, more complicated detector effects, overcoming imperfect optics) is essential for our method to extend to real data and remain superior to image-based methods like KLIP, which do not depend on accurate forward modeling of the optics.

Fig.~\ref{fig:multi_snapshot} explores the performance of our approach when incorporating multiple image measurements. 
These additional measurements were generated under different wavefront aberrations, each obtained by linearly interpolating between the two real JWST wavefront aberration measurements using different drift values. 
As the number of measurements increases, the optimization leverages the additional information to better resolve the exoplanet signal. 
The results indicate that with more image measurements, performance improves and surpasses the single-measurement case, demonstrating the potential for enhanced exoplanet detection when multiple measurements are available.

\begin{figure}[t]
\centerline{\includegraphics[width=0.98\linewidth]{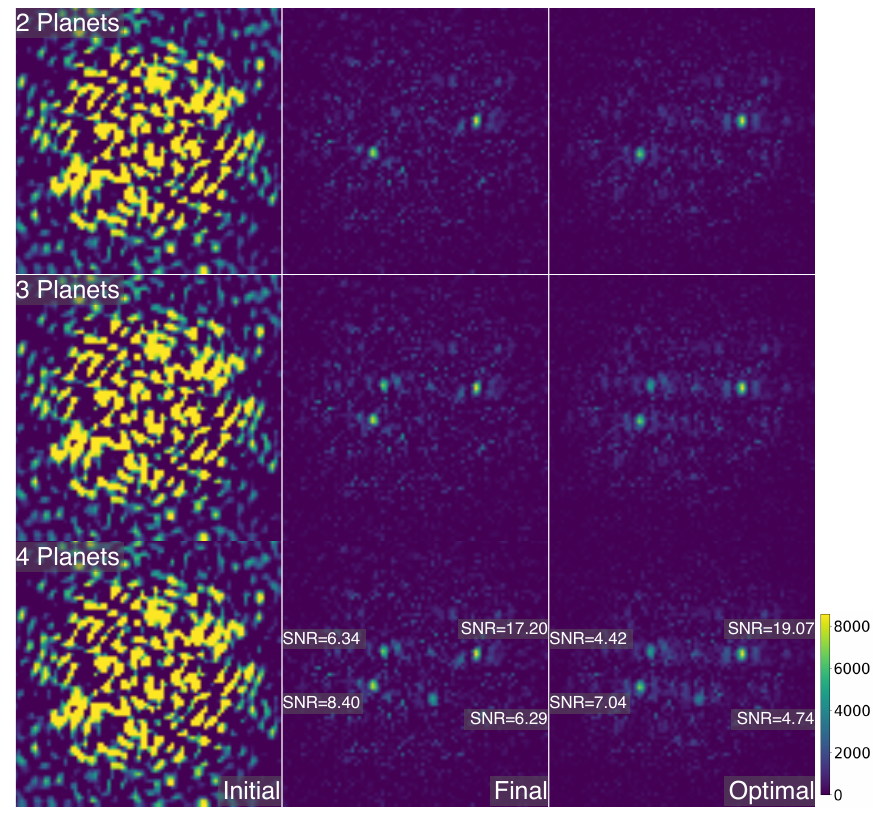}}
\caption{Performance in the presence of multiple exoplanets. The top row shows a case with two planets, the middle row with three planets, and the bottom row with four planets.
The leftmost column (\textit{Initial}) shows the initial state before optimization, the middle column (\textit{Final}) displays the subtraction results after optimization, and the rightmost column (\textit{Optimal}) illustrates the ideal scenario where the ground truth star PSF is known, revealing planet signals with optimal contrast (up to the noise limit).
Our method effectively handles multiple planets without explicitly modeling them, as it focuses solely on estimating the star PSF. This approach allows for the simultaneous revelation of any number of planets present in the system, showcasing its scalability and robustness in complex planetary scenarios.}
\label{fig:multi_planets}
\end{figure}

Our approach is agnostic to the number of planets in the system, allowing it to reveal multiple planets simultaneously if present, as demonstrated in our multi-planet simulations in Fig.~\ref{fig:multi_planets}. 
The top row shows a case with two planets, the middle with three planets, and the bottom with four planets. 
Because our method focuses solely on estimating the star PSF without explicitly modeling any planets, it naturally handles cases with any number of planets without modification. 
This is in contrast to directly modeling planet signals, which would require knowing or guessing the number of planets present and potentially adjusting the model for each case. 
This scalability is a direct consequence of our method's fundamental principle: By accurately estimating and subtracting the dominant star PSF, we allow any number of planetary signals to naturally emerge in the residual, regardless of how many are present or where they are located. 
This feature makes our method particularly robust for exploring complex planetary systems where the number of planets may not be known in advance.

\section{Sensitivity to Model Mismatch}
\label{sec:model_mismatch}
While the differentiable framework aims to accurately model the physical system, in practice, there may exist discrepancies between the simulated and the actual instrument due to imperfect knowledge or assumptions about various optical components. 
To evaluate the robustness of our approach to such model mismatch scenarios, we conducted a comprehensive sensitivity analysis by introducing controlled perturbations to the optical forward model.

Specifically, we simulated misalignments and deviations from the assumed orientation parameters for the Lyot stop, which are likely a significant source of model mismatch in practice.
These misalignments included shifts along the X-axis and Y-axis, as well as rotational misalignments, representing deviations from the nominal design and positioning of these optical elements.
For each misalignment condition, we evaluated the quality of the residual planet image obtained after subtracting the estimated star PSF from the observed science images.
For this analysis, we opted to use the peak signal-to-noise ratio (PSNR) computed over an $11\times11$ region centered on the planet, rather than the annulus SNR metric used earlier. 
While the annulus SNR is widely used in exoplanet detection, it has limitations that make it less suitable for evaluating model mismatch scenarios. 
As discussed in Section~\ref{sec:discussion}, the annulus SNR assumes a presumptive planet candidate and can be inflated even under poor subtractions with residual speckles. 
In contrast, PSNR provides a more robust measure of image quality in this particular context, allowing us to more accurately assess the impact of model mismatches on our method's performance.

\begin{figure}[t]
\centering
\includegraphics[width=0.98\linewidth]{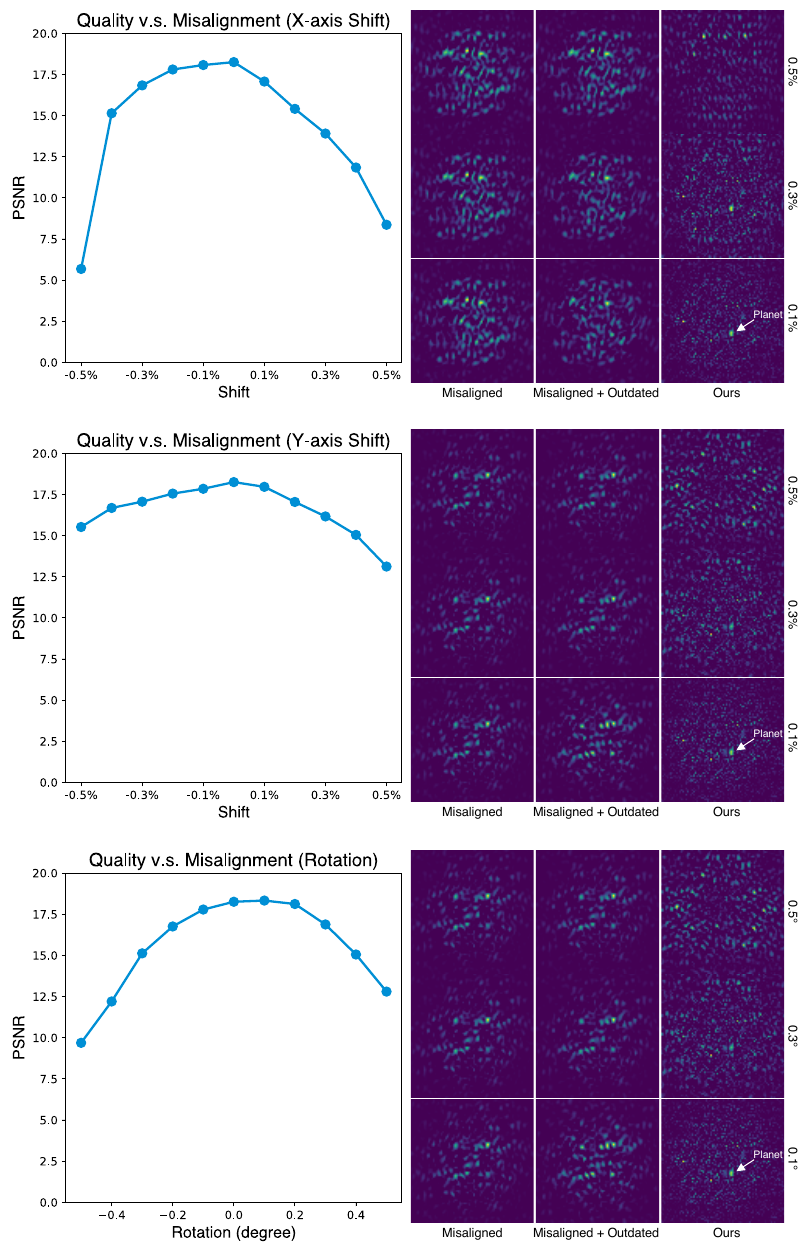}
\caption{Sensitivity analysis on forward model mismatch due to Lyot mask misalignments. We examine the impact of forward model mismatch by misspecifying the Lyot mask placement with increasing misalignments: shifts along the X and Y axes (top two rows) and rotations around the Z-axis (bottom row). 
The planet is set at a contrast level of $1\times10^{-6}$, angular separation of 0.6 arcseconds, and wavefront aberration drift of 40 hours. 
Shifts are given as percentages of Lyot mask size, and rotations are measured in degrees.
{\it Misaligned} shows subtraction results using the star PSF computed using ground truth aberration but with error due to unknown misalignments. {\it Misaligned + Outdated} shows results using the star PSF impacted by both misalignments and outdated aberrations. {\it Ours} demonstrates results obtained with our optimization pipeline, showing improved planet signal recovery across various scenarios. Results illustrate our method's robustness to moderate misalignments and forward model mismatch while also indicating the regime where severe misalignments begin to degrade performance.}
\label{fig:model_mismatch}
\end{figure}

The results of the sensitivity analysis are presented in Fig.~\ref{fig:model_mismatch}.
As expected, the results exhibit a degradation in performance as the model mismatch increases, with larger misalignments leading to a more significant drop in the image quality of the residual planet signal.
However, our differentiable pipeline maintains a robust level of performance on the residual planet image.
The graceful degradation of performance observed in these results can be attributed to the iterative optimization process. 
By leveraging gradient-based refinement, our method can partially compensate for model mismatch by adjusting the wavefront aberration estimates to better match the observed data, effectively accounting for deviations from the assumed aberrations and misalignments in the optical components.
This robustness is particularly valuable in real-world scenarios where perfect knowledge of the system parameters may not be available, or where the instrument configuration may deviate from the nominal design due to environmental factors, operational constraints, or manufacturing tolerances.
In the future, since our method is differentiable, one could attempt to parameterize these deviations and solve for them as well.

\section{Discussion}
\label{sec:discussion}
\textit{Distinction from prior art:}
Our differentiable approach marks a significant advancement in post-processing techniques for high-contrast exoplanet imaging. 
Unlike traditional methods like LOCI~\cite{Lafrenire2007ANA} and KLIP\cite{Soummer2012Detection}, our approach leverages wavefront sensing data to enhance detection performance at no added extra cost. 
Our differentiable framework can refine the star PSF using gradient-based optimization even with limited data. 
In contrast, KLIP relies on a library of reference images to construct the optimal PSF.
This need for a large library of reference images can be a significant constraint in space-based observations, where the opportunities for acquiring diverse reference data are scarce and costly. 
On the other hand, our approach relies on our ability to model the optical system accurately to reproduce the true star PSF. 
While our optical forward model will not be truly identical to that of the physical telescope, it is very likely that due to the stability of JWST, a less-than-perfect forward model will also be useful, just like the reference images taken of other stars are never identical to the target science image, and yet they have been useful to remove starlight.
The performance of our method will also likely improve with better forward model calibration and alignment to the physical telescope system.
{Compared to the early work~\cite{ygouf2013simultaneous} that proposes enhancing exoplanet detection by estimating wavefront errors, we introduce and implement a differentiable optics model that significantly improves the tractability and efficiency of the optimization problem, bringing the wave-space approach closer to practical adoption.}

\textit{More learnable variables:} 
While we only learn to adjust the optical path difference at the entrance of the telescope in this work, having a differentiable optical model may enable fitting many other parameters of interest. 
As mentioned in Section~\ref{sec:model_mismatch}, one could use this model to fit for instrument defects, such as misalignments of certain optical elements. 
Furthermore, this model could also be used to fit the precise position of the star behind the coronagraph, which could be especially important for JWST, where current methods for estimating the star position have large uncertainties that propagate and dominate the uncertainty of the relative astrometry of the planet~\cite{Carter_2022}.
The model could be extended to fit planet position and flux, but simultaneous planet detection and characterization would be more challenging, likely requiring a two-step process with detection followed by property fitting.

%\textit{Overfitting risk:} Our approach is resistant to overfitting, because it is unlikely that changing the phase of the wavefront aberration would produce a single localized point source in the image that would correspond to a planet, while simultaneously producing patterns consistent with the overwhelming star signals. For example, adding a sine pattern onto the wavefront aberrations can produce a localized point source signal at the location of a planet, but it would also produce an equally bright point source on the opposite side of the star, which would be hugely inconsistent with the observed signals. 

{\it Visual inspection vs. automated detection:} 
The standard practice for planet detection differs significantly from typical computer vision detection tasks, which involve automated detection pipelines with quantitative metrics like detection accuracy.
Exoplanet researchers generally work directly with residual images obtained from starlight subtraction, and this preference stems from several factors. 
First, experienced astronomers can leverage their domain knowledge to distinguish planetary candidates from residual speckles or artifacts. 
Second, the scarcity of confirmed exoplanet detections leads to the lack of a large, diverse dataset of confirmed exoplanet images to train and validate automated detection systems. 
Third, visual inspection allows astronomers to understand and trust the detection process fully.
Given these considerations, our method focuses on producing high-quality residual images that maximize the visibility of potential planetary signals, aligning with the prevailing practices in the astronomical community. 
Still, our approach does not preclude the future integration of automated detection methods.

{\textit{Annulus SNR:}}
The nuanced nature of the annulus SNR metric exemplifies the complexity of the problem of exoplanet detection.
The metric inherently assumes a presumptive planet candidate, as its calculation requires knowing where to extract the signal term. 
This assumption highlights a key issue: the SNR value alone does not automatically indicate the quality of the overall starlight subtraction. 
In fact, over-subtractions can inflate SNR values, and under-subtractions with numerous residual speckles might yield high SNR at certain locations and annuli. 
The issue is related to the reliance on human visual inspection in the current practice of exoplanet detection and screening, with experienced astronomers often using their visual judgment to identify promising candidates before or in conjunction with SNR calculations.
In this context, it is more appropriate to view the annulus SNR as a confidence score for a presumptive positive planet detection, which has already passed initial visual scrutiny.

\textit{Photon noise limit:}
Although our approach significantly improves the contrast and detectability of exoplanets, the final results are not perfect due to the presence of photon noise.
Even with perfect PSF subtraction, the residual image would still contain noise that can obscure faint planet signals. 
However, reaching this photon noise limit may be promising for future observations.
Unlike scenarios limited by systematic errors, where increased integration time may not improve sensitivity due to correlated noise, operating at the photon noise limit means that longer exposures will directly enhance our detection capabilities. 
This characteristic implies that investing in extended observation times will yield improved sensitivity, potentially enabling the detection of even fainter planets.

{\it Temporal flexibility of aberration estimation: }
Our method does not inherently assume a specific temporal ordering between wavefront aberration measurements and science image acquisition. 
While our simulations present a scenario where the aberration is measured earlier and the image is taken later under an evolved aberration, this is not a requirement of the approach. 
Our approach can be applied equally well when aberration measurements are taken before or after the science image, or even when measurements are available at multiple time points. 
We chose to present the ``forward-time'' scenario in our simulations as it intuitively illustrates the concept of aberration drift. 
However, the optimization formulation and process remain identical regardless of this temporal relationship. 

\red{
{\it Extension to non-monochromatic simulations:}: The assumption of monochromatic light can be relaxed in future work. 
Broadband simulations can be done by simulating several waves of different wavelengths through the optical system and performing a weighted sum to produce a broadband image. 
The weights can be obtained by combining the relative intensity of the target star being observed and the transmission of the JWST filter being simulated. 
Additionally, many known chromatic aberrations are documented in public models such as \texttt{webbPSF}, which can be incorporated in our differentiable model to make broadband simulations more accurate. 
Preliminary work using broadband simulations has shown no prohibitive increase in computing time for the optimization.}

\red{{\it Application to real JWST observations}: Future work will apply differentiable rendering to real JWST observations. While several real observations are publicly available at the Mikulski Archive for Space Telescopes (MAST), real JWST observations require extensive preprocessing (e.g., eliminating dark current and thermal background signals, identifying frames affected by cosmic rays, cleaning bad pixels). In addition, a lengthy and careful treatment of real observations is needed, since the optical aberrations might be complicated, correlated, or varying across observations. For these reasons, this paper focuses on simulated observations as a proof of concept.}

\textit{Future directions:}
Future work may relax the assumption of monochromatic point sources for more realistic broadband simulations, which are essential for applications on real astronomical data.
This extension would involve optical simulations across a range of wavelengths and combining the resulting data, to account for the chromatic instrumental response.
Moreover, the principles of our differentiable framework can be extended to other high-contrast imaging systems, both space-based and ground-based. 
For ground-based scenarios, our approach could potentially mitigate the effects of rapidly evolving atmospheric turbulence, but this would require adapting the framework to incorporate atmospheric turbulence models and handle the higher temporal frequency of aberration changes.
Future work could also explore extended sources like circumstellar disks, moving beyond the point-source planet model used in this study. 
Integration with other post-processing techniques, such as angular or spectral differential imaging, could further improve performance by combining the strengths of multiple approaches.

\section{Conclusion}
{We present a differentiable optics-based technique for exoplanet detection and demonstrate its effectiveness in enhancing detection capability over the conventional image-based approach.}
Our work connects differentiable programming with physics-based modeling in astronomical imaging, combining valuable wavefront aberration information with science camera observations. 
Our differentiable optics model allows the high-dimensional optimization problem to be tractable and efficient to solve.
By refining the star PSF via differentiable rendering through the telescope optical system, our method outperforms the state-of-the-art method KLIP without using additional reference images that KLIP and other PCA-based methods require. 
Our approach can improve exoplanet imaging efficiency by reducing observing time, conserving resources, and minimizing the need for reference observations to achieve high-quality exoplanet detections.
As we continue to refine and expand this technique, it is poised to significantly advance the search for exoplanets, accelerating the discovery of new worlds beyond our solar system and bringing us closer to answering fundamental questions about the prevalence and formation of planets in the universe.

\section*{Acknowledgment}
We thank the anonymous reviewers for their constructive feedback and suggestions. We thank Marie Ygouf, Marshall Perrin, and Laurent Pueyo for their insightful comments and help improving this paper. 
This work is in part supported by the NSF Award 2019786 (The NSF AI Institute for Artificial Intelligence and Fundamental Interactions) and the NSF CIF Award 1955864 (Occlusion and Directional Resolution in Computational Imaging).
A. L. is supported by the NSERC-Discovery Grant. 
J. J. W. and R. F.-C. are supported by STScI grants (JWST-ERS-01386 and JWST-GO-04050) under NASA contract NAS5-03127.
K. L. B. is supported by the NSF CAREER Award 2048237, the Amazon AI4Science Partnership Discovery Grant, and the Carver Mead New Adventures Fund.

\bibliographystyle{IEEEtran}
\bibliography{main.bib}

\newpage
\clearpage

\appendices
\section{Simulator design and calibration}
\label{s:appendix_A}

\subsection{Coronagraph optical elements}
The designs of the coronagraph optics were obtained from \texttt{webbPSF}~\cite{Perrin_2014}. In particular, the primary mirror design and the circular Lyot stop MASKRND were used, as well as the focal plane mask MASK335R with a pixel scale corresponding to a wavelength of $4.5$ $\mu$m. We also obtained from \texttt{webbPSF} the field of view- and wavelength-dependent OPD maps introduced by the optical instrument itself, which \texttt{webbPSF} computes from in-lab calibration testing and modeling.

\subsection{OPD measurements and processing}
To use the in-flight OPD measurements in our simulations, we perform some additional processing steps. 
The NIRCam instrument, aside from being used for coronagraphy in exoplanet imaging, is also used to perform OPD measurements through weak lenses. 
This means that the raw measurement of optical path difference has undesired contributions from the NIRCam contribution to the OPD and the field-dependent aberrations caused by the different positions of the detector when measuring the OPD and when measuring the science image of interest. 
Therefore, to obtain the OPD corresponding \textit{only} to the entrance of the instrument, we remove those contributions from the OPD measurement (also using calibration testing and modeling data through \texttt{webbPSF}). 
It is these aberration maps that are used as our ``outdated'' real measurements.

\subsection{Photometric calibration}
To compute the number of photons $n_{\text{photons}}$ at each pixel, we calibrate our contrast values to the photometry of the star $\beta$ Pictoris, which is a bright, classic direct imaging target star~\cite{Smith_1984, Lagrange_2009}. 
The calibration was done using the zero point of NIRCam in the F444W filter (184.1 Jy) and the W2 magnitude of $\beta$ Pictoris (i.e., its brightness at the wavelength $4.6$ $\mu$ m) from the WISE catalog \cite{Wright_2010} (3.182 mag).

The photometry calibration allows for the computation of the flux density of the star in units of mJy/sr corresponding to a contrast value of 1. 
Next, using calibration files from JWST, we convert to counts per second (CPS) by using the appropriate conversion factor in the F444W filter with the MASK335R focal plane mask (the one used in our simulation), which corresponds to $2.516$ mJy/sr $ = 1 $ CPS. 
Assuming a total exposure time of 2 hours, we obtain the total number of counts on each pixel. 
Using a gain of $1.82$ and a quantum efficiency of $0.8$, we obtain the number of photons per unit of contrast (contrast 1 $\sim 10^{10}$ photons).
With this simple conversion factor, our simulator then outputs images in units of photons, which allows us to also incorporate realistic photon noise to the images we wish to optimize over via a Poisson distribution. 

\red{The photon noise level at each pixel is modeled as follow: when we simulate an observation with a given photometric calibration we obtain the number of photons  $n_\text{photons}$ landing at every pixel in the detector (i.e., the intensity at each pixel), and then model the noise in each pixel with a Poisson distribution with rate $\lambda=\sqrt{n_\text{photons}}$. With the noise level at every pixel, the fundamental photon noise detection limit as a function of separation can be obtained by dividing the image in annuli centered at the center of the image, and for each annulus compute the average photon noise level (given by $\sqrt{n_\text{photons}}$ at each pixel). By establishing a detection threshold to be 5$\sigma$, this gives the minimum intensity required at that annulus to get a significant detection as a function of separation from the star. However, near the center of the coronagraph, the planet signal will also be attenuated (just like the star signal is), since it will be partially behind the focal plane mask too. To account for this, we weigh the previous photon noise limit by the focal plane mask transmission profile (that is, if at some position the focal plane mask allows only 50\% of the light to go through, a planet at that location would need to be twice as bright as the photon noise limit to still produce a 5$\sigma$ detection). This produces the fundamental noise limit shown in Figure \ref{fig:noiselimits}.}

\subsection{Read noise}
An additional source of noise is read noise in the detector. 
For our simulation, we use the READNOISE reference file\footnote{\url{https://jwst-pipeline.readthedocs.io/en/latest/jwst/references_general/readnoise_reffile.html}} from the JWST pipeline corresponding to the NIRCam instrument in the long wavelength channel (where the filter F444W is) with the SUB320A335R subarray (i.e., the region of the detector corresponding to the coronagraphic mask used, MASK335R). 
This file provides an estimate of the mean level of noise between two frames at each pixel (which is nearly uniform in the entire region of the detector) in units of counts from in-flight measurements. 
For our simulations, we take the average value of the noise level reported in this file divided by $\sqrt{2}$ (since we want the noise in a single frame, not the noise between two frames) as the amplitude of independent Gaussian noise at each pixel. 

This estimate is an effective over-estimation of the level of read noise. 
The reason is that JWTS's infrared detectors use non-destructive reads of the detector for up-the-ramp readouts\footnote{\url{https://jwst-docs.stsci.edu/understanding-exposure-times}} to estimate photon flux from photon counts. 
The read noise level reported in reference files corresponds to the noise between those non-destructive reads (Correlated Double Sampling, or CDS, noise), and the up-the-ramp readouts reduce the effects of this noise in the flux images by fitting a line to read counts. 
The reduction factor of effective read noise from up-the-ramp readouts is approximately the square root of the number of samples, which can be on the order of tens to thousands in real observations.

For simplicity, we use the CDS noise (divided by $\sqrt{2}$, see above) as a total read noise in the final image, which is an effective over-estimation of the noise level. 
However, read noise is much less significant than photon noise, especially in the region of the image of interest for exoplanet detection. 
For reference, the average number of photons at a separation of $0.7\pm0.2$ arcseconds is $\sim 10^6$, so an associated level of Poisson noise of $\sim 1000$ photons. 
On the other hand, the level of read noise (uniform across the image) from CDS is $\sim 42$ photons. Therefore, while our effective read noise is slightly over-estimated, photon noise is still overwhelmingly dominant. We include this read noise for completeness, but it will not contribute significantly to the performance limit of our method. 

\red{Part of what makes photon noise dominant is the low temperature at which JWST operates, together with the relatively long exposures (on the order of tens of minutes to a few hours for total exposure time) of almost static targets. While effective read noise from up-the-ramp fitting is reduced with more and more samples, the level of photon noise increases with more and more samples, thus resulting in noise dominated by Poisson photon noise.}

\begin{figure*}[t]
\centerline{\includegraphics[width=0.98\linewidth]{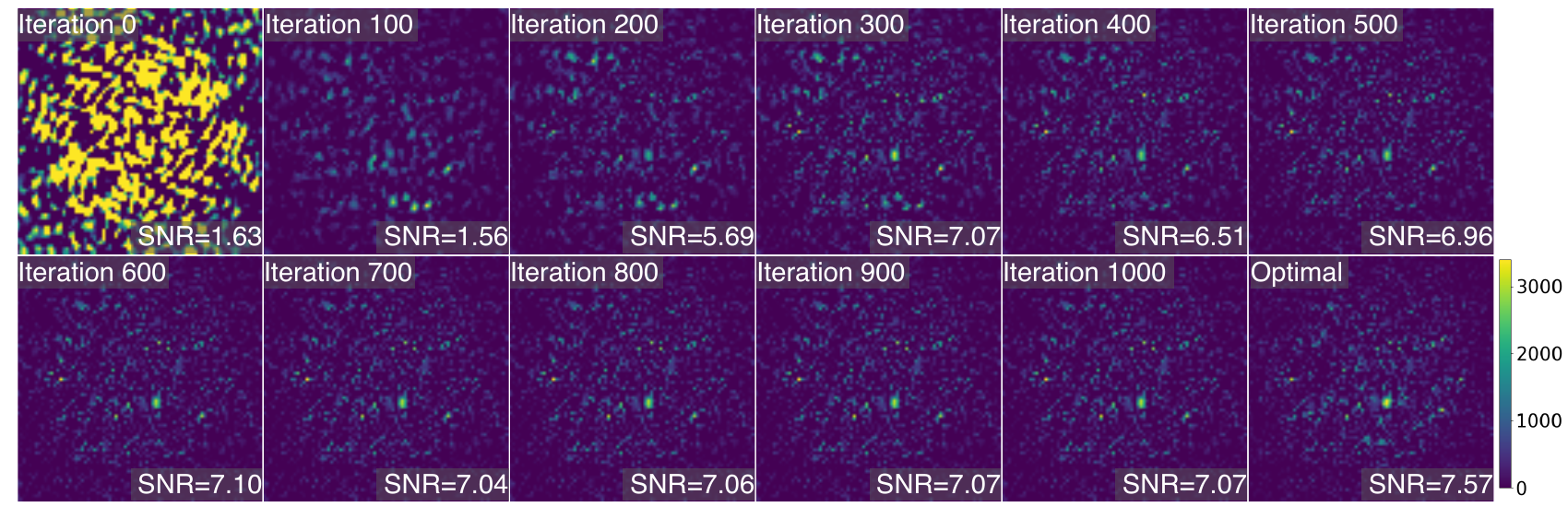}}
\caption{Progression of the subtraction result over the course of optimization. 
This simulation assumes a planet-star contrast level of $1\times10^{-6}$, angular separation of 0.6 arcseconds, and 80 hours of wavefront aberration drift.
The estimated wavefront aberration map $\hat{\phi}$ is iteratively improved using gradient descent, starting from an outdated initial measurement (outdated after 80 hours of drift). 
As the optimization progresses, the corresponding starlight subtraction using the refined star PSF increasingly suppresses speckles and reveals the planet signal more effectively. The {\it Optimal} image shows the ideal result obtained by subtracting the ground truth star PSF, with imperfections due to measurement noise.}
\label{fig:optim_progress}
\end{figure*}

\begin{figure}[t]
\centerline{\includegraphics[width=0.98\linewidth]{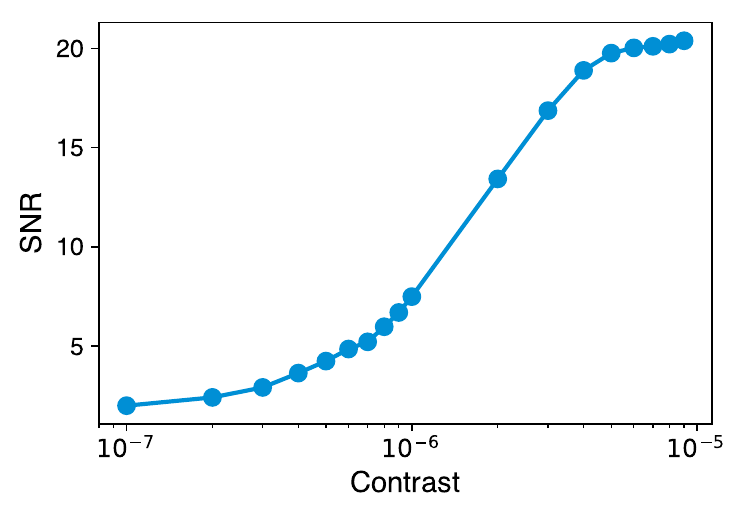}}
\caption{Performance across various contrast levels. Fixing the planet at a 0.6 arcseconds separation from the star and assuming a wavefront aberration drift of 40 hours, we show SNR values that measure the planet recovery quality for different contrast levels. The results here provide details on the gradual degradation in performance as the planet becomes fainter, given the same separation and wavefront aberration drift levels.} 
\label{fig:performance_curve_contrast}
\end{figure}
\begin{figure}
\centerline{\includegraphics[width=0.98\linewidth]{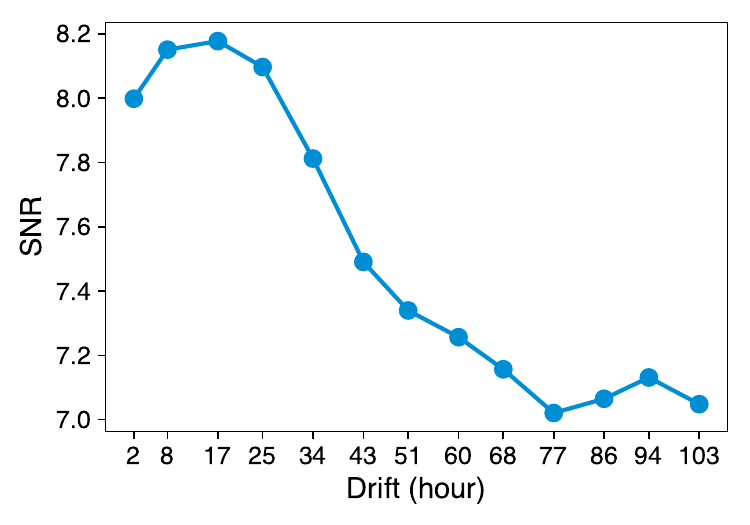}}
\caption{Performance across various wavefront aberration drift levels. Fixing the planet at a 0.6 arcseconds separation from the star and assuming a planet-star contrast of $10^{-6}$, we show SNR values that measure the recovery quality for varying drift levels.
The consistent performance improvement across all levels demonstrates the method's ability to recover planet signals even when starting from significantly inaccurate initial wavefront aberration estimates.}
\label{fig:performance_curve_drift}
\end{figure}
\begin{figure}
\centerline{\includegraphics[width=0.98\linewidth]{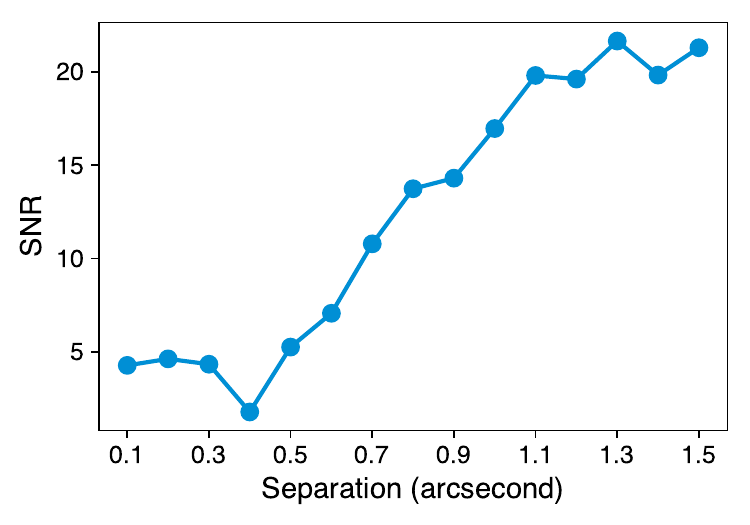}}
\caption{Performance across various angular separations between planet and star. We assume a planet-star contrast of $10^{-6}$ and a wavefront aberration drift of 40 hours. The results demonstrate that our method consistently improves the quality of the recovered planet PSF at intermediate to large angular separations. As expected, performance degrades at angular separations less than 0.5 arcseconds, where planets become fainter due to occultation by the coronagraphic instruments designed to block starlight.}
\label{fig:performance_curve_separation}
\end{figure}

\section{Additional results}
\label{s:appendix_B}

Fig.~\ref{fig:optim_progress} provides details into the optimization process.
The figure displays intermediate results for a single simulated case with a contrast of $8\times10^{-6}$ and wavefront aberration drift of around 80 hours.
Starting from an initial, outdated wavefront aberration estimate, our method iteratively improves this estimate using gradient descent. 
At each step, the updated $\hat{\phi}$ is used to render an estimated image $\hat{y}$, which is compared to the observed image $y$ to compute the loss. 
As the optimization progresses, the rendered images increasingly resemble the observations, and the final, refined star PSF enables a more effective starlight subtraction.

Fig.~\ref{fig:performance_curve_contrast}, \ref{fig:performance_curve_drift},~\ref{fig:performance_curve_separation} provide a quantitative assessment of the performance of our approach in enhancing the quality of the recovered planet signal. 
We evaluate the annulus SNR of the planet after subtracting the estimated star PSF from the observation.
In Fig.~\ref{fig:performance_curve_contrast}, we show the performance across different star-planet contrast levels.
As the contrast level becomes more challenging, the SNR decreases, reflecting increasing difficulty in detecting the planet signal. 
Fig.~\ref{fig:performance_curve_drift} examines the performance under different levels of drift. 
Our approach achieves a robust performance across all drift levels.
Fig.~\ref{fig:performance_curve_separation} shows the performance under different angular separation levels between the planet and the star.
As expected, lower angular distances are considerably more challenging, but our method achieves a graceful degradation in performance as the angular distances increase.

\section{Simulated PSF library for KLIP comparison}
\label{s:appendix_C}

The reference library used for KLIP in Fig.~\ref{fig:klip_comparison} was collected in a manner similar to the simulation of the true target image.
While our simulations followed the properties of $\beta$ Pictoris, our reference star simulations for KLIP followed the star $\alpha$ Pictoris, which has been recently used in practice as the reference star when directly imaging $\beta$ Pictoris b with JWST~\cite{Krammerer_2024}.
The photometric calibration was done in an identical way to $\beta$ Pictoris, except using the W2 magnitude of $\alpha$ Pictoris (0.979 mag) from the WISE catalog \cite{Wright_2010} and an exposure time for each image of 400 seconds. Nine different images following a 9-POINT-CIRCLE dither pattern were generated.  Photon noise and read noise were also added to these reference images. 

Recent observations of directly-imaged exoplanets with JWST have made the reference star observations relatively close in time to their respective target star observations (e.g., ~\cite{Carter_2022,Krammerer_2024}), primarily to minimize the variability of wavefront errors between target and reference. To mimic this difference, for every experiment where KLIP was run for a target with a given wavefront error drift, the reference images were generated assuming the same wavefront error drift time plus 3 hours. So reference images have extremely similar, but not identical, optical path differences.

One important limitation of reference star differential imaging is the difficulty of having the target star and the reference star perfectly aligned, since small differences in the star position behind the coronagraph translate into important differences in the PSF. This is, in fact, the very reason that the images of the reference star are taken at several slightly different (sub-pixel) alignment angles, hoping that one of the alignments will match very closely the alignment of the target star. On-sky testing has found that the errors in target alignment can be up to 20 milliarcseconds~\cite{Girard_2022}, with good alignment between target and reference being under 5 mas (e.g., the F200W filter in~\cite{Krammerer_2024}), and exceptionally bad alignment near 50 mas (e.g., the F444W filter in~\cite{Krammerer_2024}). For reference, the width of a pixel in the long wavelength channel of JWST is $\sim 63$ mas. To account for this difficulty, in our 9-POINT-CIRCLE pattern to generate reference images, the central point is shifted from the real location of the target star by 10 mas ($\sim 1/6$ of a pixel) to the right, an offset well within the expected misalignment levels. Since these reference images are very close to the size of the star PSF (and to the size of the target image), this shift will cause an effective under-subtraction on the left edges of some images. In real JWST data, the images are much larger than the star PSF, so the relative shift does not typically cause the patterns seen in Figure \ref{fig:klip_comparison}.

To contextualize the comparison of our method with PCA-based methods, we note that as more observations are made with JWST, there is an ongoing effort to build larger PSF libraries, thus performing KLIP not only on a single reference star but on a large amount of them. 
While the same limitations will still apply (e.g., differences in OPD drift, star alignment, spectral types, instrument systematics, the drawback of dedicating observing time to these reference stars, etc.), a big and diverse enough collection of reference PSFs can, in principle, significantly improve the performance of PCA-based methods in PSF subtraction.
\iffalse
\section{Biography Section}
If you have an EPS/PDF photo (graphicx package needed), extra braces are
 needed around the contents of the optional argument to biography to prevent
 the LaTeX parser from getting confused when it sees the complicated
 $\backslash${\tt{includegraphics}} command within an optional argument. (You can create
 your own custom macro containing the $\backslash${\tt{includegraphics}} command to make things
 simpler here.)
\vspace{11pt}

\iffalse
\bf{If you include a photo:}\vspace{-33pt}
\begin{IEEEbiography}[{\includegraphics[width=1in,height=1.25in,clip,keepaspectratio]{fig1}}]{Michael Shell}
Use $\backslash${\tt{begin\{IEEEbiography\}}} and then for the 1st argument use $\backslash${\tt{includegraphics}} to declare and link the author photo.
Use the author name as the 3rd argument followed by the biography text.
\end{IEEEbiography}

\vspace{11pt}

\bf{If you will not include a photo:}\vspace{-33pt}
\begin{IEEEbiographynophoto}{John Doe}
Use $\backslash${\tt{begin\{IEEEbiographynophoto\}}} and the author name as the argument followed by the biography text.
\end{IEEEbiographynophoto}
\fi

\fi

\vfill

\end{document}